\documentclass[final,authoryear,12pt]{elsarticle}
\usepackage{geometry} 
\usepackage{amssymb}
\journal{Icarus}

\begin{document}

\begin{frontmatter}



\title{Titan's past and future: 3D modeling of a pure nitrogen atmosphere and geological implications}


\author[label1,label2]{Benjamin Charnay}
\author[label1]{Fran\c cois Forget}
\address[label1]{Laboratoire de M\'et\'eorologie Dynamique, CNRS,
Universit\'e P\&M Curie (UPMC), Paris, France}
\address[label2]{Virtual Planetary Laboratory, University of Washington, Seattle, Washington 98195, USA.}
\author[label3]{Gabriel Tobie}
\address[label3]{Laboratoire de Plan\'etologie et G\'eodynamique, UMR-CNRS 6112, Universit\'e de Nantes, Nantes, France.}
\author[label4]{Christophe Sotin}
\address[label4]{Jet Propulsion Laboratory, California Institute of Technology, Pasadena, California, USA.}
\author[label5]{Robin Wordsworth}
\address[label5]{Department of the Geophysical Sciences, University of Chicago, Chicago, Illinois, USA.}

\begin{abstract}

Several clues indicate that Titan's atmosphere has been depleted in methane during some period of its history, possibly as recently as 0.5-1 billion years ago. It could also happen in the future. Under these conditions, the atmosphere becomes only composed of nitrogen with a range of temperature and pressure allowing liquid or solid nitrogen to condense. 
Here, we explore these exotic climates throughout Titan's history with a 3D Global Climate Model (GCM) including the nitrogen cycle and the radiative effect of nitrogen clouds.
We show that for the last billion years, only small polar nitrogen lakes should have formed. 
Yet, before 1 Ga, a significant part of the atmosphere could have condensed, forming deep nitrogen polar seas, which could have flowed and flooded the equatorial regions. Alternatively, nitrogen could be frozen on the surface like on Triton, but this would require an initial surface albedo higher than 0.65 at 4 Ga. Such a state could be stable even today if nitrogen ice albedo is higher than this value.
According to our model, nitrogen flows and rain may have been efficient to erode the surface. Thus, we can speculate that a paleo-nitrogen cycle may explain the erosion and the age of Titan's surface, and may have produced some of the present valley networks and shorelines. Moreover, by diffusion of liquid nitrogen in the crust, a paleo-nitrogen cycle could be responsible of the flattening of the polar regions and be at the origin of the methane outgassing on Titan.

\end{abstract}


\end{frontmatter}

% \linenumbers

\section{Introduction}
Titan's atmosphere is thick ($\sim$ 1.47 bar), essentially composed of N$_2$ (more than 95 $\%$) and methane ($\sim$5 $\%$ close to the surface and $\sim$1.5 $\%$ above the tropopause) \cite[]{niemann10}. Methane photodissociation generates a complex chemistry, leading to the formation of H$_2$, organic molecules and haze.
Currently, Titan's surface temperature is around 93 K \cite[]{jennings11}. It is controlled by a greenhouse effect dominated by collision-induced absorption (CIA) of N$_2$-N$_2$, CH$_4$-N$_2$ and H$_2$-N$_2$, and by the absorption of sunlight by CH$_4$ and haze in the upper atmosphere, generating an anti-greenhouse effect \citep{mckay91}.

 The inventory of total carbon (atmospheric methane, lakes, sand dunes,...) present on the surface of Titan seems far smaller (2-3 orders of magnitude) than the amount estimated to have been produced throughout Titan's history, as estimated from the present rate of methane photolysis \citep{lorenz08, sotin12}. Hence, this carbon might have a recent origin.
\cite{tobie06} suggested that Titan's atmospheric methane originated from episodic outgassing, which released methane from clathrates, starting approximately 0.5-1 billion years ago. This is consistent with the dating derived from isotopic analysis of C$^{12}$/C$^{13}$, which provides an upper limit of no more than 470 Ma for methane outgassing \citep{mandt12}, as well as with the time required for the formation of the dunes, estimated to range between 50 and 630 Ma \citep{sotin12}. Before this outgassing, Titan's atmosphere might therefore have been depleted in methane and its photochemical products. In such conditions, the greenhouse effect was limited to the CIA of N$_2$-N$_2$, and the atmosphere was colder and could have condensed \citep{lorenz97}, forming  liquid nitrogen on the surface. 

This state could also occur in the future. Indeed, atmospheric CH$_4$ has a lifetime of about 20 million years \citep{krasnopolsky09}. 
Thus, if it is not resupplied, it disappears, together with all its photoproducts. In such a case, haze particles are no more produced and H$_2$ molecules escape from the atmosphere. The atmosphere is therefore exclusively composed of nitrogen after approximately 10 million years. 
Currently, no source of methane, able to maintain the present-day level in Titan's atmosphere has been identified. There might be a subsurface source of methane explaining the detection of possible tropical lakes \citep{griffith12}. Half of the methane could be resupplied by ethane diffusion in polar clathrates \citep{choukroun12}, but it would not be sufficient. Titan could therefore end up in the liquid nitrogen state within a few million years.

\cite{lorenz97} studied the surface temperature and pressure under such conditions with a 1D model. They found that, with a faint Sun and a high albedo (higher than 0.5), the atmosphere might have undergone a collapse leading to a Triton-like frozen state, with a thin atmosphere. Their model did not include clouds, assuming that they would quickly fall to the ground. 
In this paper, we study the climates and atmospheric collapse of a pure nitrogen atmosphere with a 3D Global Climate Model (GCM) during Titan's history. The model incorporates cloud formation and their radiative effect. In section 2, we describe the model and the assumptions on the nitrogen cycle. We then  analyze the radiative impact of clouds, theoretically and with the GCM. We study the nitrogen cycle and atmospheric collapse for past and future climates. Finally, we discuss the implications of a paleo-nitrogen cycle on the surface erosion and shape, and methane outgassing on Titan.

\section{Method}

Simulations were performed using a new type of GCM, the Generic LMDZ, specifically developed for exoplanet and paleoclimate studies \citep{wordsworth11,wordsworth13,forget13,leconte13b,charnay13}. 
The radiative scheme was based on the correlated-k model, with the absorption data calculated directly from high resolution spectra computed by a line-by-line model from the HITRAN 2008 database \citep{rothman09}. 
Rayleigh scattering by N$_2$ is included, using the method described in \cite{hansen74}, and using the \cite{toon89} scheme to compute the radiative transfer. 
The N$_2$-N$_2$ continuum from the HITRAN database, fundamental for this study, was included.
We used 16 spectral bands in the thermal infrared and 18 at solar wavelengths. 

The nitrogen cycle and cloud modeling is based on physical principles. We used the same method as CO$_2$ clouds on early Mars in \cite{forget13}. The nitrogen condensation is assumed to occur when atmospheric temperature drops below the saturation temperature from \cite{armstrong54}. Local mean N$_2$ cloud particle sizes are determined from the amount of condensed nitrogen and the number density of cloud condensation nuclei (CCN) by:

 \begin{equation}
r=\left( \frac{4 q}{3 N_{c} \pi \rho_{N_2}}\right)^{1/3}
 \end{equation}
with r the mean cloud particle radius, $q$ the mass mixing ratio of condensed nitrogen, $N_c$ the number density of CCN per mass unit of air, and $\rho_{N_2}$ the volumic mass of condensed nitrogen.

The values used for the number density of CCN are discussed in the next section. We did not make any distinction between liquid and icy particles for the radiative transfer. Single scattering properties were calculated considering spherical particle Mie theory with the optical properties of nitrogen ice of \cite{quirico96}. As for Rayleigh scattering, the radiative transfer for nitrogen cloud in the GCM uses the \cite{toon89} scheme.
Liquid and icy particles were assumed to sediment according to Stokes law \citep{forget99} and evaporate during their fall. 
No coalescence of liquid droplets is taken into account because the number density of CCN in our simulations is too small to allow this process (see next section).

No ground infiltration for liquid nitrogen is taken into account. Therefore, liquid nitrogen from precipitation or surface condensation is conserved on the surface.
Surface condensation (evaporation) of N$_2$ occurs when the surface temperature goes below (above) the saturation temperature. It is calculated from energy conservation principles, using a latent heat for N$_2$ of 198 kJ/kg. When condensation or evaporation occurs on the ground or in the atmosphere, the atmospheric pressure is adjusted in consequence.
In this study, all simulations were initiated with a surface pressure of 1.47 bar, similar to present-day pressure on Titan. The lack of methane should slightly reduce the pressure, but the change is too small to affect our results.
We used a surface emissivity of 1 and a thermal inertia for the ground of 400 J s$^{-1/2}$m$^{-2}$K$^{-1}$, a value estimated by \cite{tokano05}. We have run the model with a higher inertia of 2000  J s$^{-1/2}$m$^{-2}$K$^{-1}$, and did not notice any change on the mean surface temperature at any latitude. We also implemented a diffusion scheme for the liquid nitrogen at the surface, representing slow surface flows. We used a diffusivity of 100 m$^2$s$^{-1}$, equal to the one used by \cite{schneider12} for liquid methane. Most of our simulations have been run without such a diffusivity, but we discuss its impact.

One of the main parameters in this study is the surface albedo. For simplicity, we 
assumed a constant value over all Titan. We neglected a change of the surface albedo by liquid nitrogen. Indeed, liquid nitrogen is not radiatively active at visible wavelengths and so its impact on the albedo should be limited.
The present surface albedo should be around 0.2-0.3 \citep{schroder08}, pretty small compared to other saturnian moons (i.e. $\sim$0.4-0.6 \citep{howett10}). Enceladus has a very high albedo around 0.95, but it is a particular case with a permanent resurfacing produced by a geological activity. The Galilean moons have a similar or higher albedo (around 0.2 for Callisto, 0.3 for Ganymede, and 0.5 for Io and Europa). Triton, which is covered by ices of water, nitrogen and methane, has a high albedo around 0.7. 
Before the methane outgassing the surface albedo on Titan was certainly higher than today because of the lower amount of dark organic material on the surface. Thus, it was likely between 0.3 and 0.5, closer to Bond albedo values of the other saturnian and Galilean moons. In this study, most of the simulations were run with different values of albedo, varying from 0.2 to 0.5. Yet, we took an albedo of 0.3 as reference, in particular to estimate the impact of nitrogen clouds (next section).

The simulations were run with a mean solar insolation at top of Titan's atmosphere of 3.77 W/m$^2$ for the present-day Sun. For paleo-climates we use the \cite{gough81} law to calculate past insolation. We used the present-day values for astronomical parameters (i.e. Saturn's distance to the Sun, Saturn's eccentricity and obliquity). We neglect the fact that  Saturn's distance to the Sun could have been different at 4 Ga, before the Late Heavy Bombardment \citep{tsiganis05}.

In the early Solar System, Saturn was warmer. According to the thermal evolution model from \cite{leconte13a}, Saturn's effective temperature was between 130 and 200 K at 4 Ga (compared to $\sim$96 K today), corresponding to an additional infrared warming of Titan between 0.01 and 0.06 W/m$^2$. Yet, this remains small compared to the solar flux on Titan (i.e. 3.8 W/m$^2$ for the present Sun and 2.8 W/m$^2$ at 4 Ga) and we therefore neglected it.

\section{Effects of nitrogen clouds}

\begin{figure}
\begin{center}
\includegraphics[width=20pc]{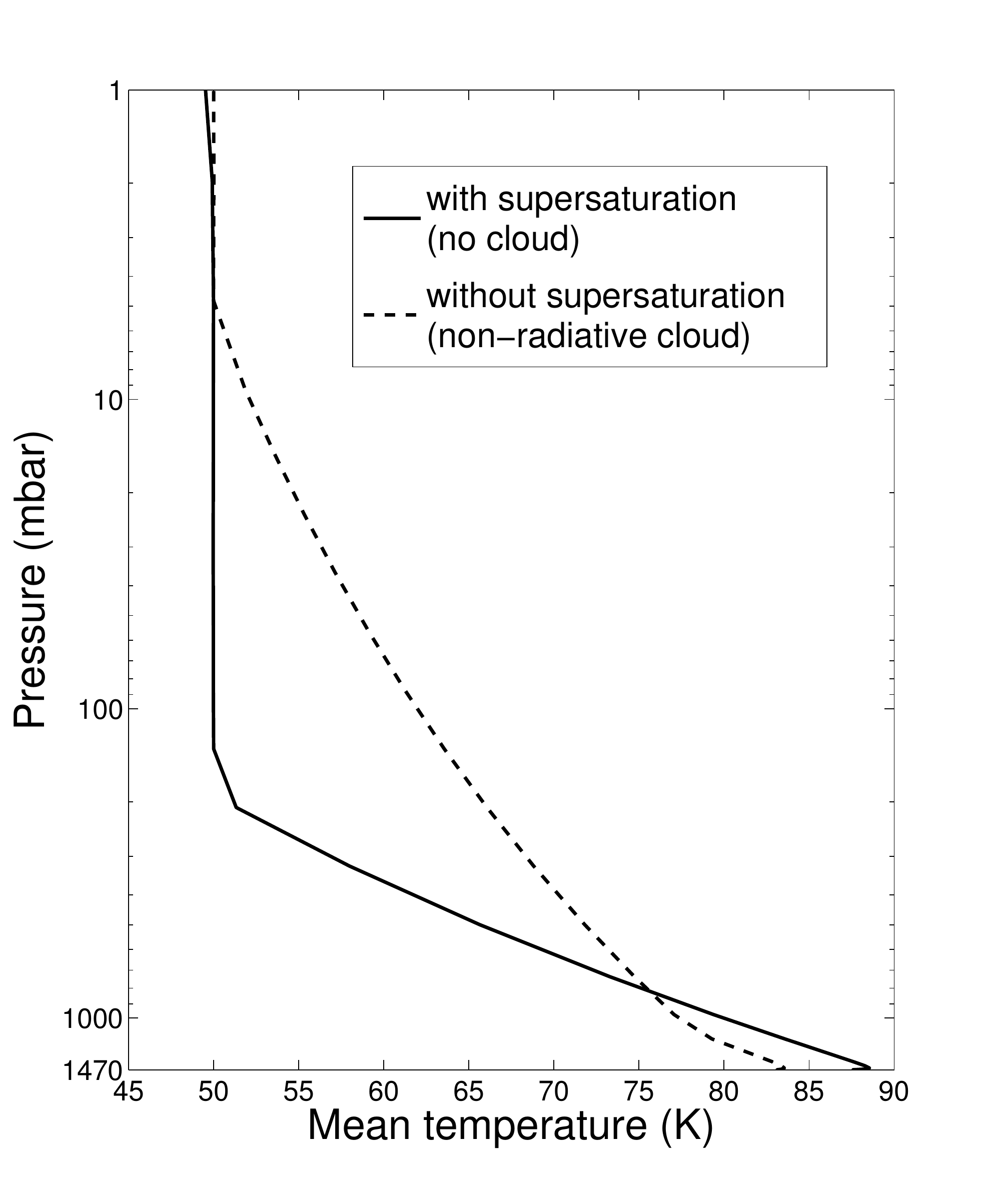}
\caption{Profiles of mean temperature for a supersaturated atmosphere with no N$_2$ condensation (solid line) and a saturated atmosphere with non-radiative cloud (dashed line). Simulations were performed with the present solar irradiance and a surface albedo of 0.3. }
\label{fig_temp}
\end{center}
\end{figure}

We first ran the GCM for a pure nitrogen atmosphere with non-radiative cloud to analyse the possibility for nitrogen cloud formation. Two cases were considered, when the atmosphere can condense, producing clouds, and when nitrogen condensates cannot nucleate, even at high supersaturation. Figure \ref{fig_temp} shows the mean vertical temperature profile for both cases with a surface albedo of 0.3. Condensation occurs between 1000 and 6 mb (i.e. 6 km and 70 km).
The surface temperature is 4.5 K higher with no nucleation. In both cases, the temperature in the stratosphere becomes close to 50 K, as compared to 180 K for the present stratosphere. In the middle troposphere, the temperature with no nucleation is lower, inducing a stronger greenhouse effect. The maximum difference is reached at 200 mb (i.e. at $\sim$30 km), with a value of 15 K. This corresponds to a huge supersaturation of 2500 $\%$. Such a level of supersaturation could not be maintained. Condensation would quickly occur on any CCN. Moreover, homogeneous nucleation would probably happen. We therefore conclude that a fully supersaturated atmosphere cannot exist. Moderate supersaturation may happen (e.g. up to 50 $\%$ corresponding to less than 2 K of difference), producing a small change of the surface temperature. Nevertheless, for simplicity, we will assume in following analyses that the atmosphere condenses when saturation is reached (no supersaturation).
A sudden change from the supersaturated profile to the saturated profile would correspond to a condensation of around 2000 kg/m$^2$ (a depth of 2.5 m of liquid nitrogen).

Now, we investigate the radiative effect of nitrogen clouds. Liquid or solid nitrogen is very transparent at most thermal infrared wavelengths. Therefore, unlike water clouds on Earth, nitrogen clouds do not absorb thermal radiation. They can scatter the upcoming thermal flux if the cloud particles are large enough, and induce a "scattering greenhouse effect", as described by \cite{forget97} for CO$_2$ clouds.

 If we assume a fully scattering nitrogen cloud layer in the atmosphere and a surface emissivity of 1, by neglecting the Rayleigh scattering and the greenhouse effect of N$_2$, we have the next three equations:
 \begin{equation}
F_{s  \uparrow}^{vis}=A_s F_{s \downarrow}^{vis}
 \end{equation}
 \begin{equation}
F_{s  \downarrow}^{vis}=(1-A_c^{vis})F_0+A_c^{vis} F_{s \uparrow}^{vis}
 \end{equation}
\begin{equation}
\sigma T_s^4=(1-A_s^{vis})F_{s  \downarrow}^{vis}+\sigma T_s^4 A_c^{ir}
 \end{equation}

with $F_0$ the solar flux at the top of the atmosphere, $F_{s  \uparrow}^{vis}$ and $F_{s  \downarrow}^{vis}$ the upward and downward visible fluxes at the surface,  $\sigma$ the constant of Stefan-Boltzmann, $T_s$ the surface temperature, $A_s$ the surface albedo, $A_c^{vis}$ the visible albedo of clouds, $A_c^{ir}$ the infrared albedo of clouds.

Using these three equations, we can express the warming produced by nitrogen clouds:

 \begin{equation}
\Delta T_s=\left(\left[\frac{1}{(1-A_s A_c^{vis})} \frac{(1-A_c^{vis})}{(1-A_c^{ir})}\right]^{1/4}-1\right) T_s^{0}
 \end{equation}
where $T_s^{0}=(\frac{(1-A_s)F_0}{\sigma})^{1/4}$ is the surface temperature without cloud. Thus, the warming effect increases with the surface albedo.

In reality, the calculation is far more complicated because of the Rayleigh scattering and the greenhouse effect of nitrogen.
Yet, if the clouds are high in the atmosphere (a reasonable approximation in our simulations), the warming effect of clouds is approximately given by equation (1), replacing $A_s$ by the planetary albedo (without cloud) and $T_s^{0}$ by the effective temperature (without cloud).

\begin{figure}
\begin{center}
\includegraphics[width=20pc]{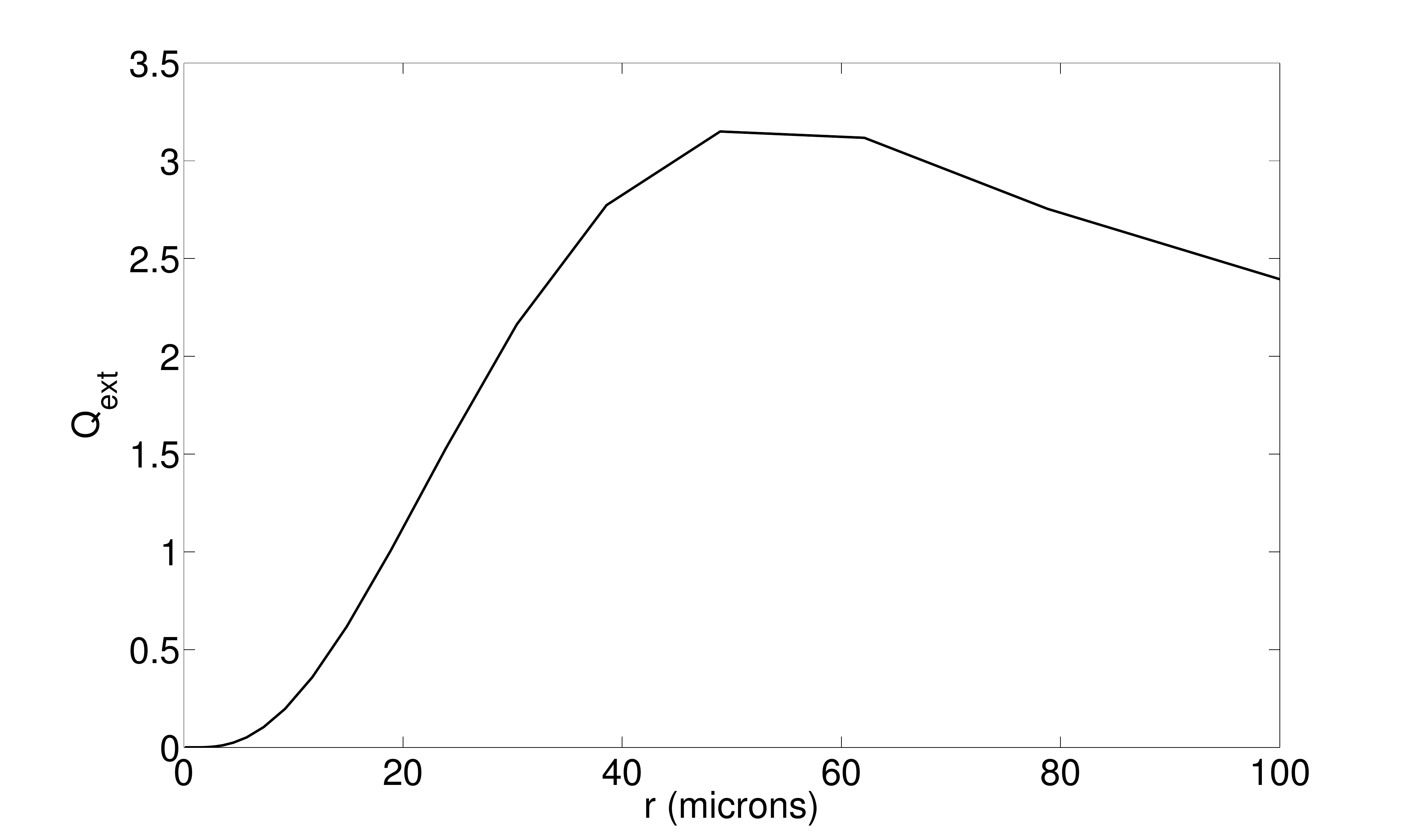}
\caption{Dependence on radius of the efficiency factor of scattering at 38 microns.}
\label{qext}
\end{center}
\end{figure}

The albedo of a cloud is linked to the optical depth given by:
\begin{equation}
\tau=\frac{3}{4}\frac{Q_{ext} w}{\rho r_e}
\end{equation}
with $Q_{ext}$ the single scattering extinction coefficient, $r_e$ the effective radius of the cloud particles,  $\rho$ the volumic mass of the particles and $w$ the mass column of the cloud.
$Q_e$ is typically equal to 2 at visible wavelengths for the cloud particle radii we used (10-100 $\mu$m).
At infrared wavelengths, $Q_e$ strongly depends on the radius.
For a surface temperature of 77 K (what we expect for a surface albedo of 0.5), the peak of the thermal emission is at 38 $\mu m$. Figure \ref{qext} shows the extinction coefficient dependence on the radius at this wavelength. For a radius above 30 $\mu m$, $Q_e$ is larger than 2. Under these conditions, nitrogen clouds lead to warming.  When the radius becomes smaller than 30 $\mu m$, $Q_e$ rapidly decreases and the cloud becomes transparent to infrared radiation. The visible albedo effect of the clouds therefore dominates, resulting in surface cooling.

\begin{figure}
\begin{center}
\includegraphics[width=20pc]{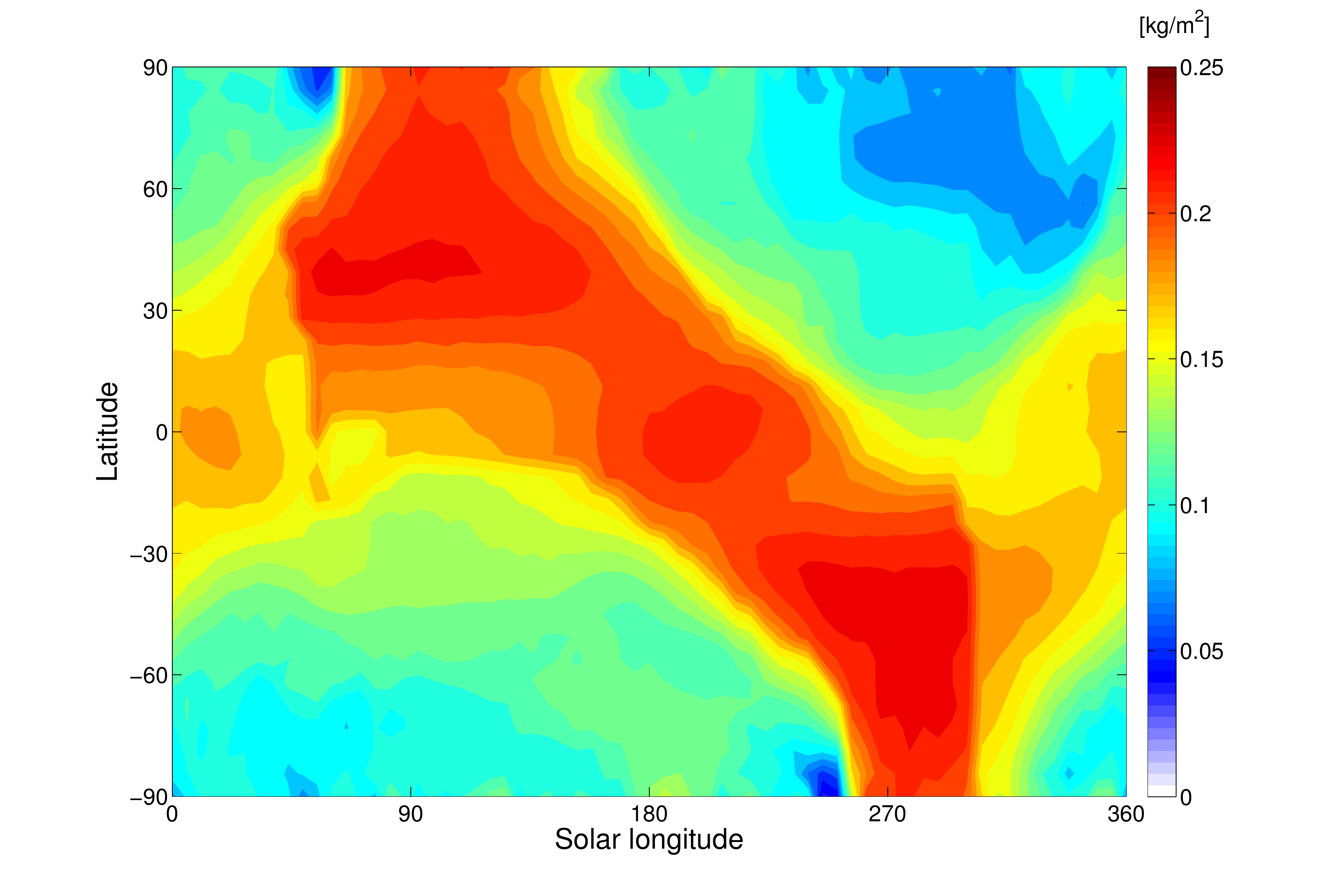}\\
\includegraphics[width=20pc]{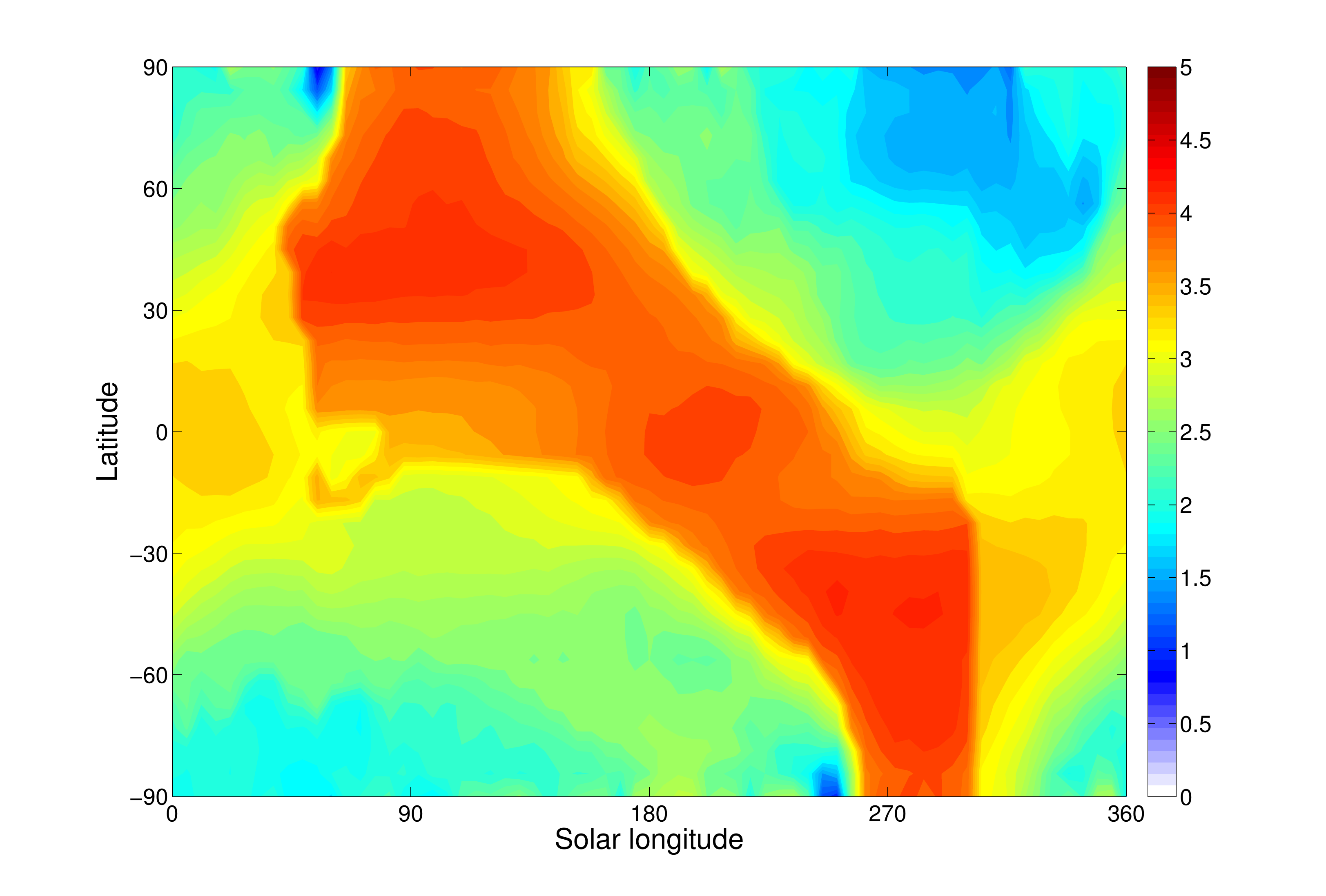}
\caption{Zonally averaged column of condensed nitrogen (top) and visible optical depth (down) depending on the solar longitude. Simulation performed with the present insolation, a surface albedo of 0.3,  $N_c$=10$^3$ particles/kg and with radiative clouds.}
\label{fig_cloud}
\end{center}
\end{figure}

In our model, the radius of cloud particles is controlled by N$_c$, the number of CCN per mass of air. This parameter has therefore a strong impact on the surface temperature. 
Table \ref{table1} gives the surface temperature, the surface pressure and the planetary albedo for the different cases considered.  Clouds can lead to a warming up to +7.8 K. This warming increases with the number density of CCN up to N$_c$$\sim$10$^5$ kg$^{-1}$.
We noticed a strong cooling for N$_c$ $\geq$ $10^6$. The clouds become transparent to infrared radiation and the planetary albedo is so high that a dramatic collapse of the atmosphere happens. Such a density of CCN is of the same order of magnitude as on present Titan ($\sim$10$^7$-10$^8$ kg$^{-1}$) \citep{tomasko05}, where CCN originate essentially from organic haze. A cooling effect by nitrogen clouds is therefore impossible for an atmosphere depleted in methane.
The amount of CCN should also be small compared to Earth ($\sim$10$^5$-10$^{10}$ kg$^{-1}$) and Mars ($\sim$10$^5$ kg$^{-1}$) \cite[]{forget13}, because of the absence of CCN source and of a very weak dust transport. Nevertheless, micro-meteorites and icy dust from Hyperion may act as CCN \citep{banaszkiewicz97, krivov01}.
For paleo- and future climate simulations, we used the case with non-radiative cloud as a reference. This corresponds to a very small density of CCN. We also ran simulations with radiative clouds and N$_c$=10$^3$ kg$^{-1}$. We consider this as an upper limit for the density of CCN, corresponding to a flux of CCN by precipitation of around 10$^3$ particles/m$^2$/s. With this value, cloud particle radii are up to 100 $\mu m$. Figure \ref{fig_cloud} shows the zonally averaged column of condensed nitrogen (liquid or solid) and the optical depth depending on the season, for N$_c$=10$^3$ kg$^{-1}$, the present insolation, a surface albedo of 0.3 and with radiative clouds. Clouds are present at any latitude and at any season. The longitudinal variations are small. Then, there is a 100$\%$ cloud covering.

\section{Simulations of paleo- and future climates}

\begin{figure}
\begin{center}
\noindent\includegraphics[width=17pc]{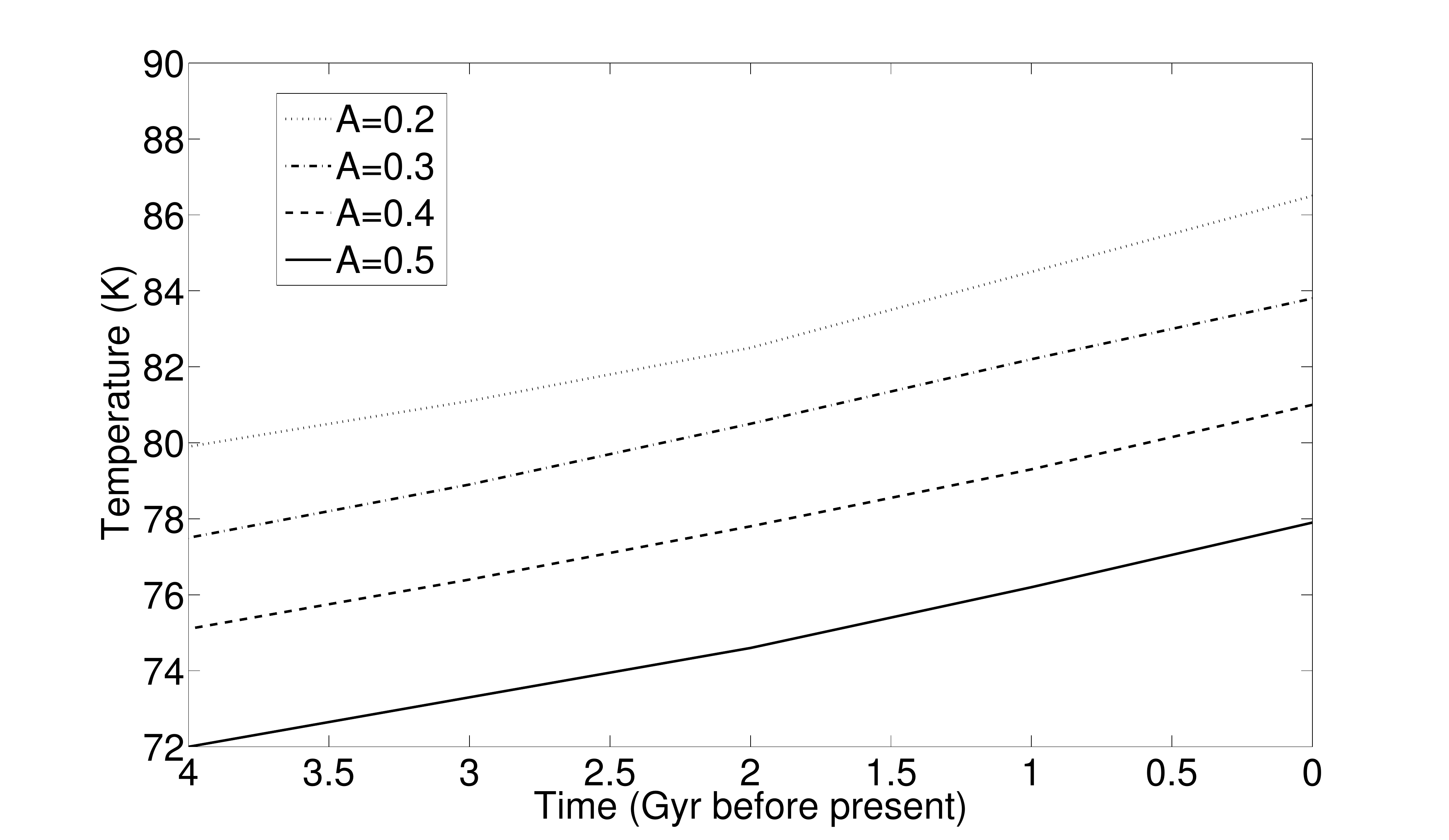}
\noindent\includegraphics[width=17pc]{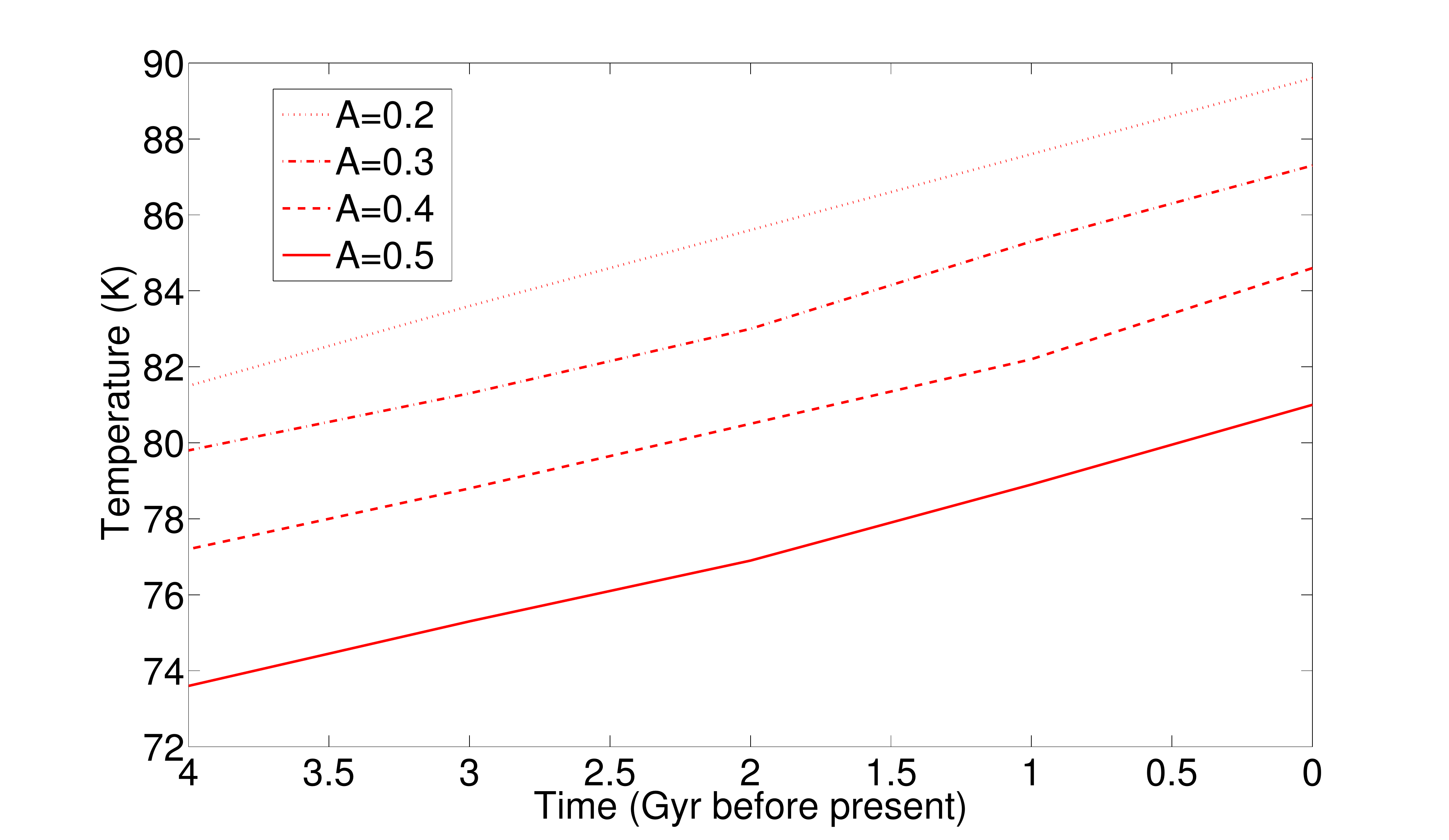}\\
\noindent\includegraphics[width=17pc]{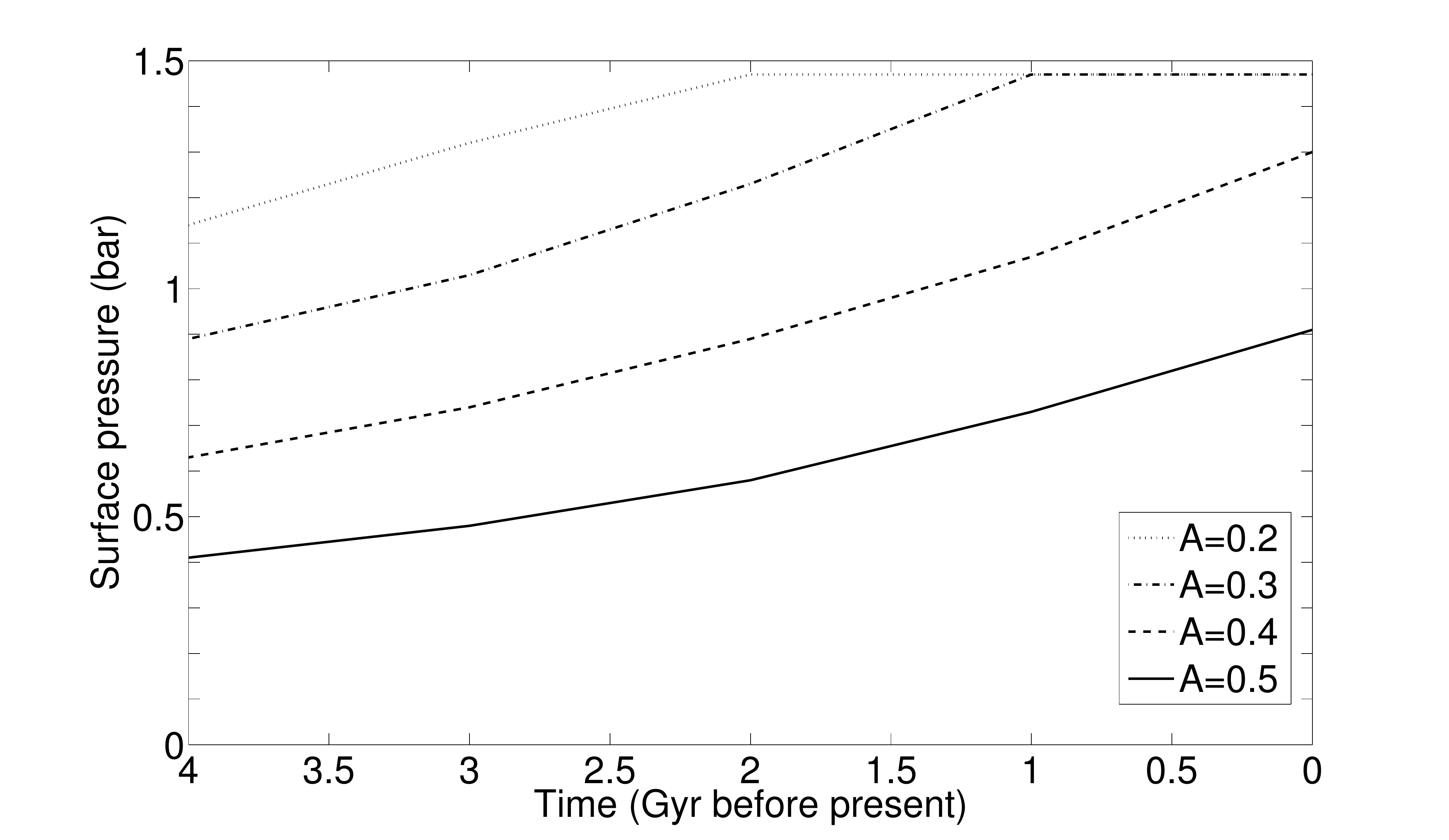}
\noindent\includegraphics[width=17pc]{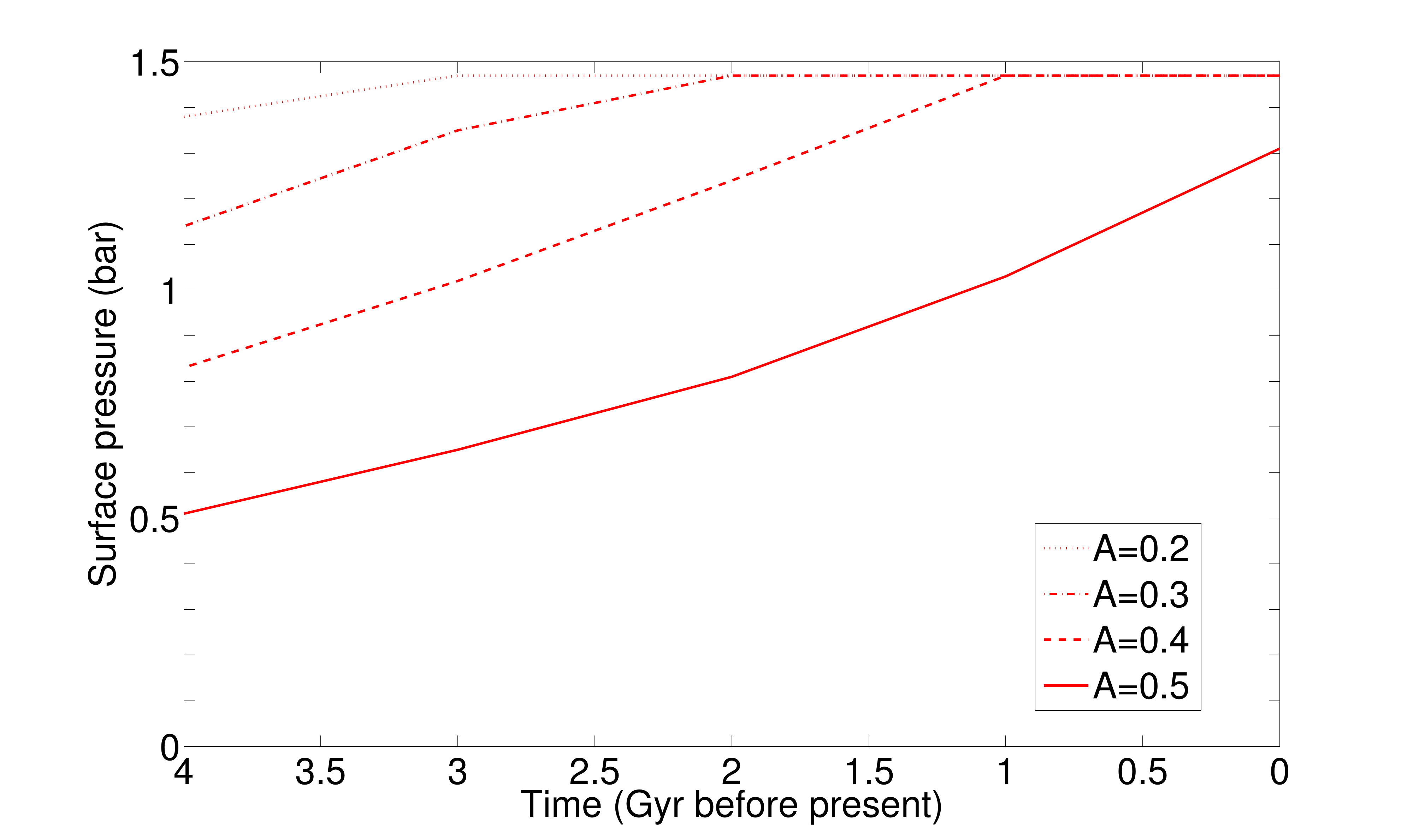}
\caption{Surface temperature (top) and pressure (down) during Titan's history with surface albedo of 0.2, 0.3, 0.4 or 0.5 (dotted lines, dashed-dotted lines, dashed lines and fill lines respectively) with non-radiative clouds (left) or with radiative clouds (right).}
\label{fig_ps}
\end{center}
\end{figure}

To explore paleo- and future climates on Titan, we ran simulations with radiative and non-radiative clouds, with a global surface albedo varying from 0.2 to 0.5, and  at 4 Ga, 3 Ga, 2 Ga, 1 Ga and nowadays, corresponding respectively to a mean solar insolation at top of the atmosphere of 2.79, 2.99, 3.21, 3.47, 3.77 W/m$^2$ \citep{gough81}. 
Figure \ref{fig_ps} shows the surface temperature and pressure for these different conditions. The radiative clouds produce a scattering greenhouse warming of 2-4 K. Thus, the impact of radiative clouds remains limited and does not change the main conclusions of our analysis. We also ran a few simulations with a higher number density of CCN (i.e. N$_c$=10$^5$ kg$^{-1}$). In this case, the climate is a little warmer than with N$_c$=10$^3$ kg$^{-1}$ for the present-day insolation. Yet, with a weaker Sun, the climate with N$_c$=10$^5$ kg$^{-1}$  is a little colder than with N$_c$=10$^3$ kg$^{-1}$, because there are more lower clouds, which have a cooling effect. We can therefore limit our study to the cases with N$_c$=10$^3$ kg$^{-1}$, with radiative or non-radiative clouds.
We now detail these simulations for future, "recent" (1 Ga or less) and early Titan.

\subsection{Future climates}
We consider here the case where Titan loses its methane in the next millions of years. This corresponds to the simulations with the present solar constant and a surface albedo of 0.2, close to the present one. The mean temperature is 86.5 K with non-radiative clouds and 89.5 K with radiative clouds. With non-radiative clouds, a small amount of liquid nitrogen (up to a depth of 1.5 m) appears in the polar regions during winter, where the surface temperature decreases to 80.7 K (see Figure \ref{fig_n2_ts}). With radiative clouds, no liquid nitrogen is maintained on the surface. Nitrogen clouds still condense in the atmosphere, yet precipitation evaporate before reaching the surface. Thus, in the future, Titan without methane should be surrounded by thin nitrogen clouds, but its surface should remain very dry with no liquid nitrogen or only episodically at the poles.

\subsection{Recent climates}
We consider here Titan's climate before the outgassing of methane, 0.5-1 Gyrs ago.
This corresponds to simulations with the solar constant at 1 Ga with or without radiative clouds. We explored climates with surface albedo from 0.2 to 0.5, higher than today due to the lower amount of dark organic material on the surface. 

\begin{figure}
\begin{center}
 \hskip -1.5cm
\noindent\includegraphics[width=14pc]{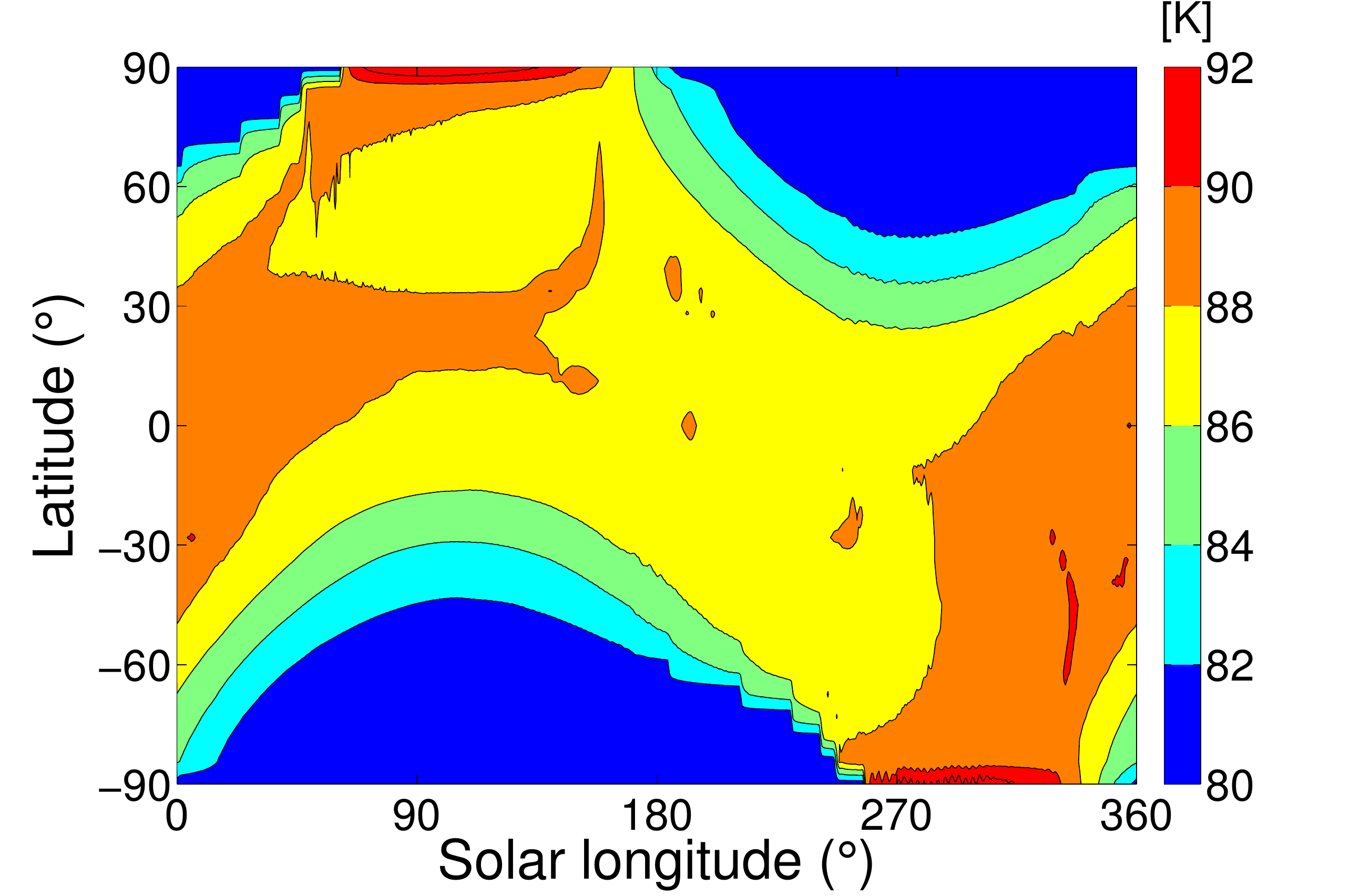} \hskip -0.5cm
\noindent\includegraphics[width=14pc]{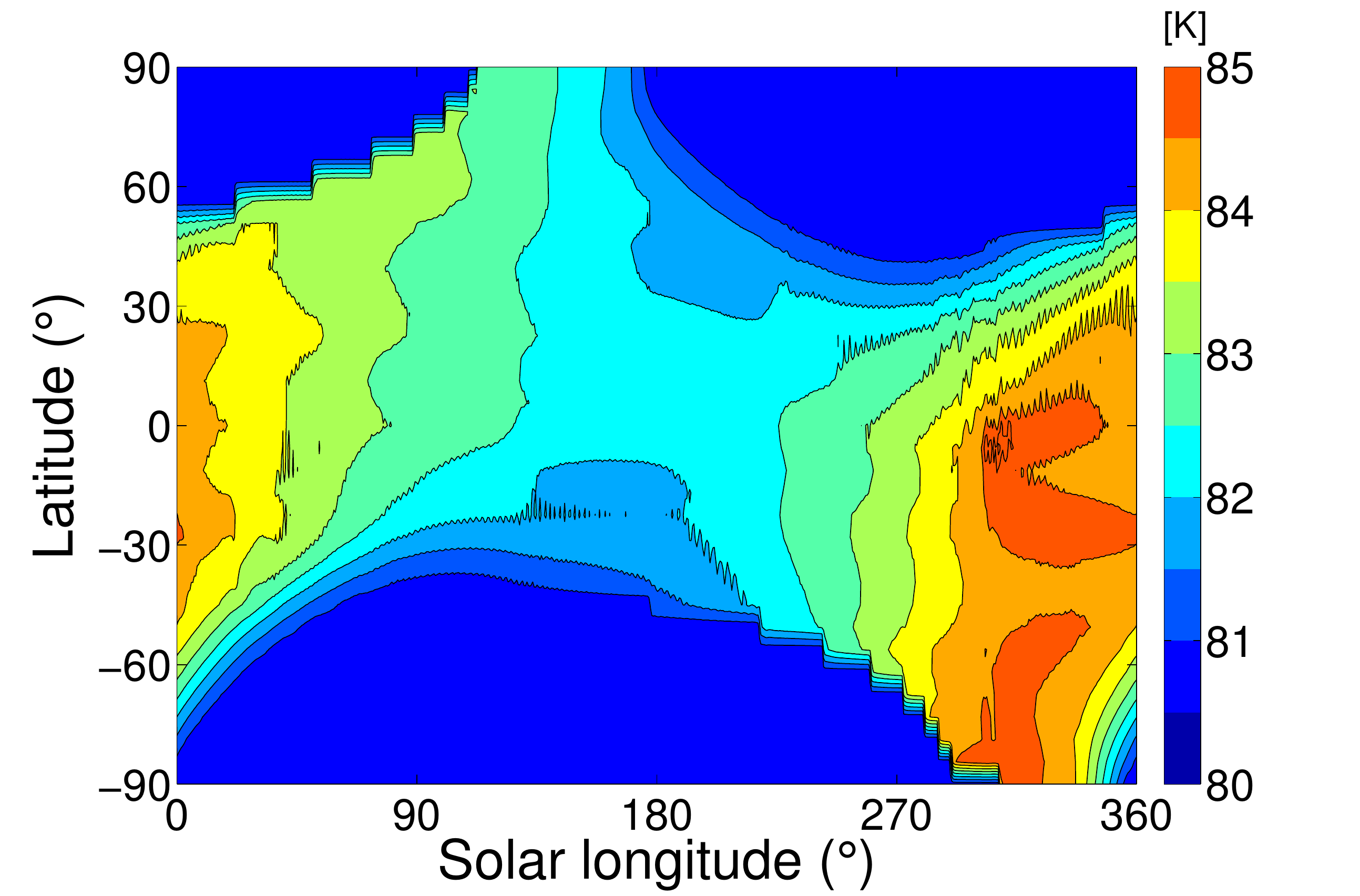} \hskip -0.5cm
\noindent\includegraphics[width=14pc]{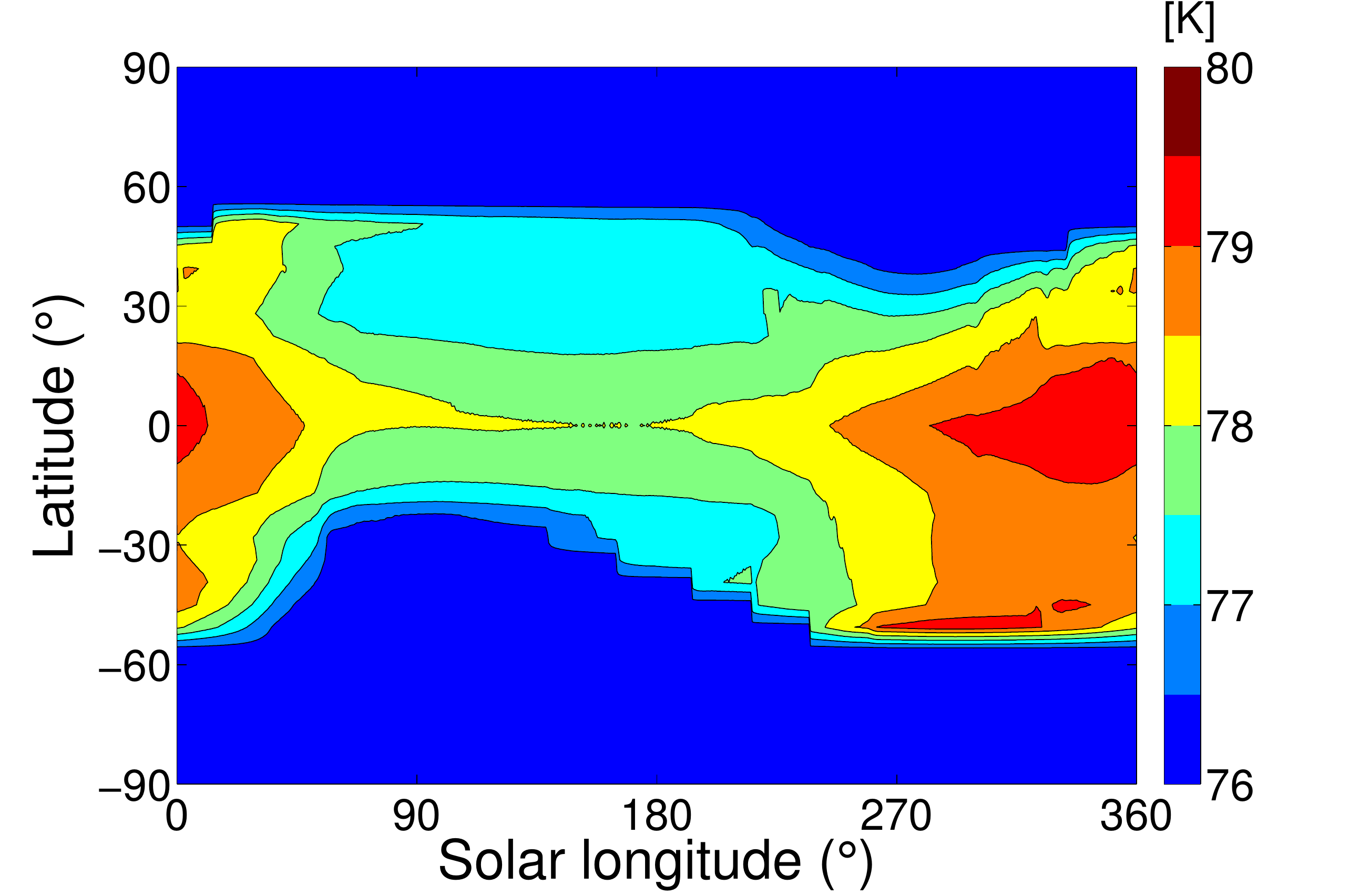} 

\hskip -1.5cm
\noindent\includegraphics[width=14pc]{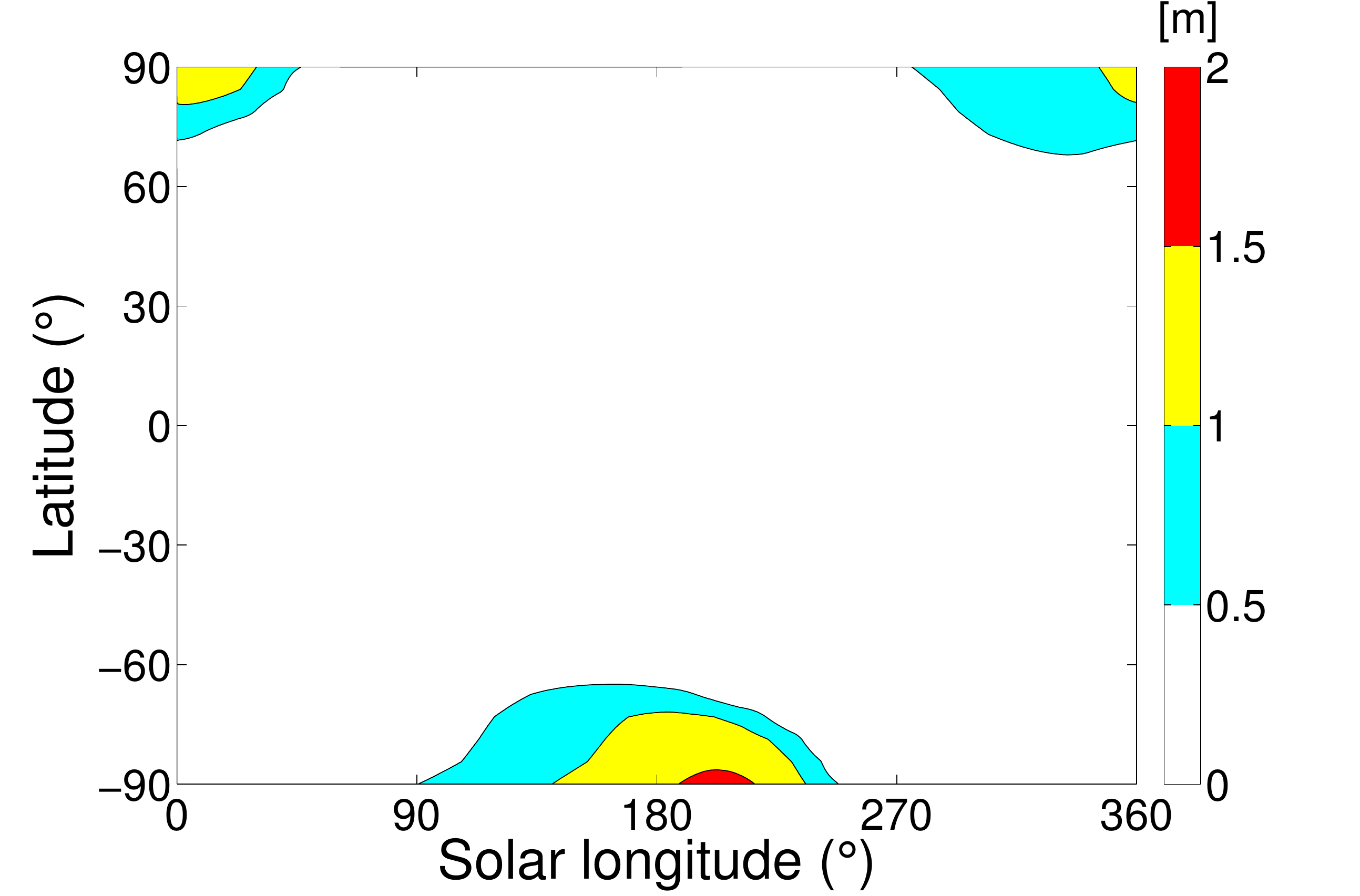} \hskip -0.5cm
\noindent\includegraphics[width=14pc]{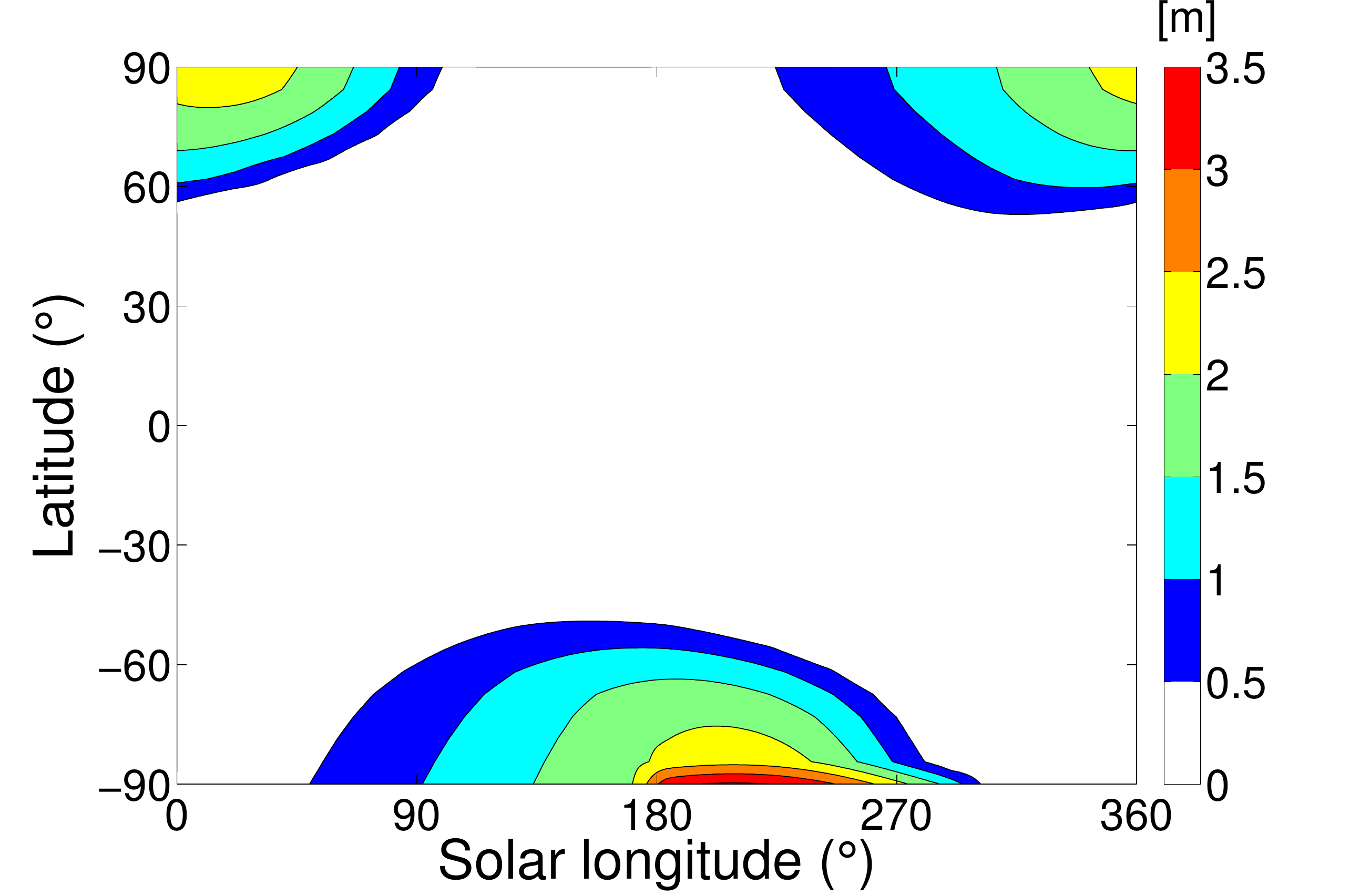} \hskip -0.5cm
\noindent\includegraphics[width=14pc]{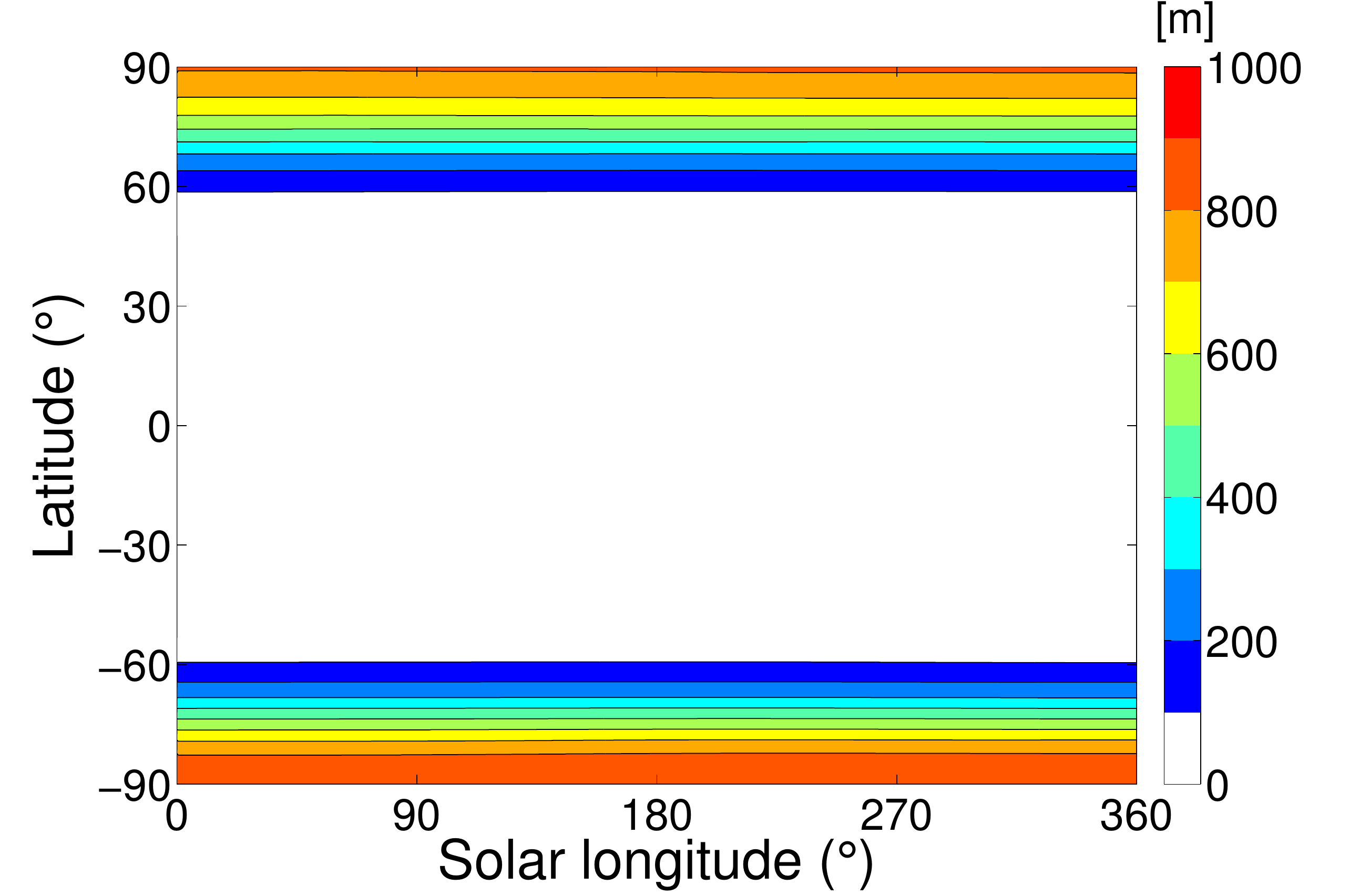} 
\caption{
Zonally averaged surface temperature (top) and depth of liquid nitrogen (in m) on the surface (down), depending on latitude and solar longitude. Simulations have been done at present with an albedo at 0.2 (left panels), at 1 Ga with an albedo at 0.3 (middle panels) and at 4 Ga with an albedo at 0.3. All simulations have been done with non-radiative clouds.}
\label{fig_n2_ts}
\end{center}
\end{figure}

For a low albedo (e.g. 0.2 and 0.3), liquid nitrogen appears in the polar regions during winter. This episodic liquid nitrogen corresponds to a typical average depth of 0.5-1 m for latitudes higher than 70$^{\circ}$, rising up to 3 m in winter (see Fig. \ref{fig_n2_ts}).  This case, is similar to simulations for future climates (discussed in the previous section), but liquid nitrogen is present for a longer time each year.
Because of Titan's topography and craters, liquid nitrogen lakes should be smaller but deeper than in our simulations without topography. They might be able to resist full evaporation during the summer, like current hydrocarbons lakes and seas on the North pole.
Hence, if Titan was depleted in methane during the last billion of years but with a low albedo (0.3 or less), the condensation should have been limited with essentially polar nitrogen lakes.

For higher values of albedo (0.4-0.5 with non-radiative clouds or 0.5 with radiative clouds), nitrogen remains permanently condensed at high latitude, forming permanent large seas. 
A decrease by 0.1 bar of the surface pressure corresponds to an average depth of around 9 m of liquid nitrogen. 
Since modeled surface pressure can be as low as 0.7 bar, large amounts of nitrogen could have condensed (up to 0.75 bar). These cases with permanent deep seas are similar to those described in the next paragraph for early climates.

\begin{figure}
\begin{center}
\includegraphics[width=20pc]{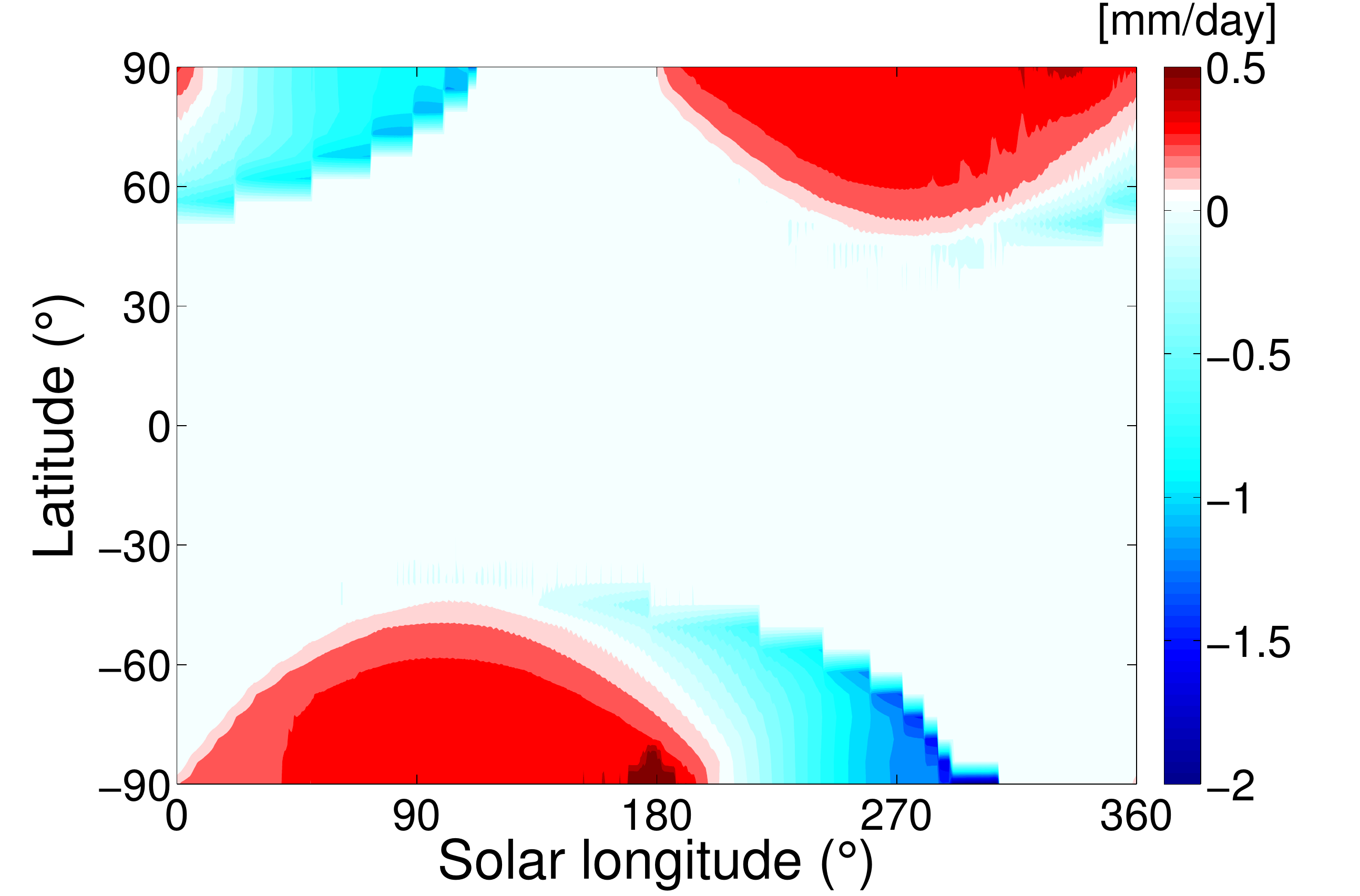}
\caption{Zonally averaged accumulation (precipitation and surface condensation) or evaporation of liquid nitrogen depending on latitude and solar longitude. Simulation performed at 1 Ga with the surface albedo at 0.3 and with non-radiative clouds.}
\label{fig_rain_cond}
\end{center}
\end{figure}

In all our simulations, we notice an asymmetry, in the distribution of condensed nitrogen in the polar regions. More liquid nitrogen is maintained in the south (see Fig. \ref{fig_n2_ts}). This is due to the colder winters in the southern pole because of Saturn's orbit eccentricity. This is opposite to the current distribution of methane lakes \citep{aharonson09} explained by longer northern summers, associated with more precipitation in the northern polar regions \citep{schneider12}. Such a difference with the current methane cycle is expected since the nitrogen cycle operates differently, with all the atmosphere in equilibrium with the coldest region. In the polar regions, our model produces an accumulation in winter (see Fig. \ref{fig_n2_ts} and Fig. \ref{fig_rain_cond}). This is the exact opposite of the model of \cite{schneider12}.
However, this asymmetry has to reverse over long periods of time (i.e. higher than 45000 years).

\subsection{Early climates}
By reducing the solar constant, a significant change in surface pressure occurs for any surface albedo and with radiative or non-radiative clouds. 

If we assume that Titan's albedo was likely between 0.3 and 0.5, closer to Bond albedo values of the other saturnian moons, Titan's atmosphere would have undergone a partial collapsing during its early history with a surface pressure between 0.4 and 0.9 bar at 4 Ga (see Fig. \ref{fig_ps}). 
Large amounts of nitrogen could have condensed during Titan's early history, forming deep seas at high latitudes (see Fig. \ref{fig_n2_ts}), which could have flooded low latitudes (see next section).

We can also imagine a case where the surface albedo was higher than 0.5 on early Titan. Table 2 shows the mean surface temperature and pressure at 4 Ga with non-radiative cloud and for a surface albedo varying from 0.2 to 0.8. According to our model, an albedo higher than 0.65 is required at 4 Ga to trigger nitrogen freezing at the surface, corresponding to a surface pressure below the triple point of N$_2$ (i.e. 0.125 bar). With nitrogen ice on the surface, the albedo could rise to $\sim$0.7-0.8, as on Triton, stabilizing this frozen state.
According to our model, an albedo of 0.8 is sufficient to maintain a frozen state even today.

\section{Wet past climates}
The climates described in the previous section are dry with liquid nitrogen only at high latitudes. Under these conditions, precipitation at low and mid latitudes evaporates before having reached the surface.
In this section, we study the possibility for a wet climate (i.e. with liquid nitrogen at low latitude) during Titan's past and we analyze the differences between such a climate and the dry climates previously described.

\subsection{Liquid nitrogen flows}
\begin{figure}
\begin{center}
\includegraphics[width=20pc]{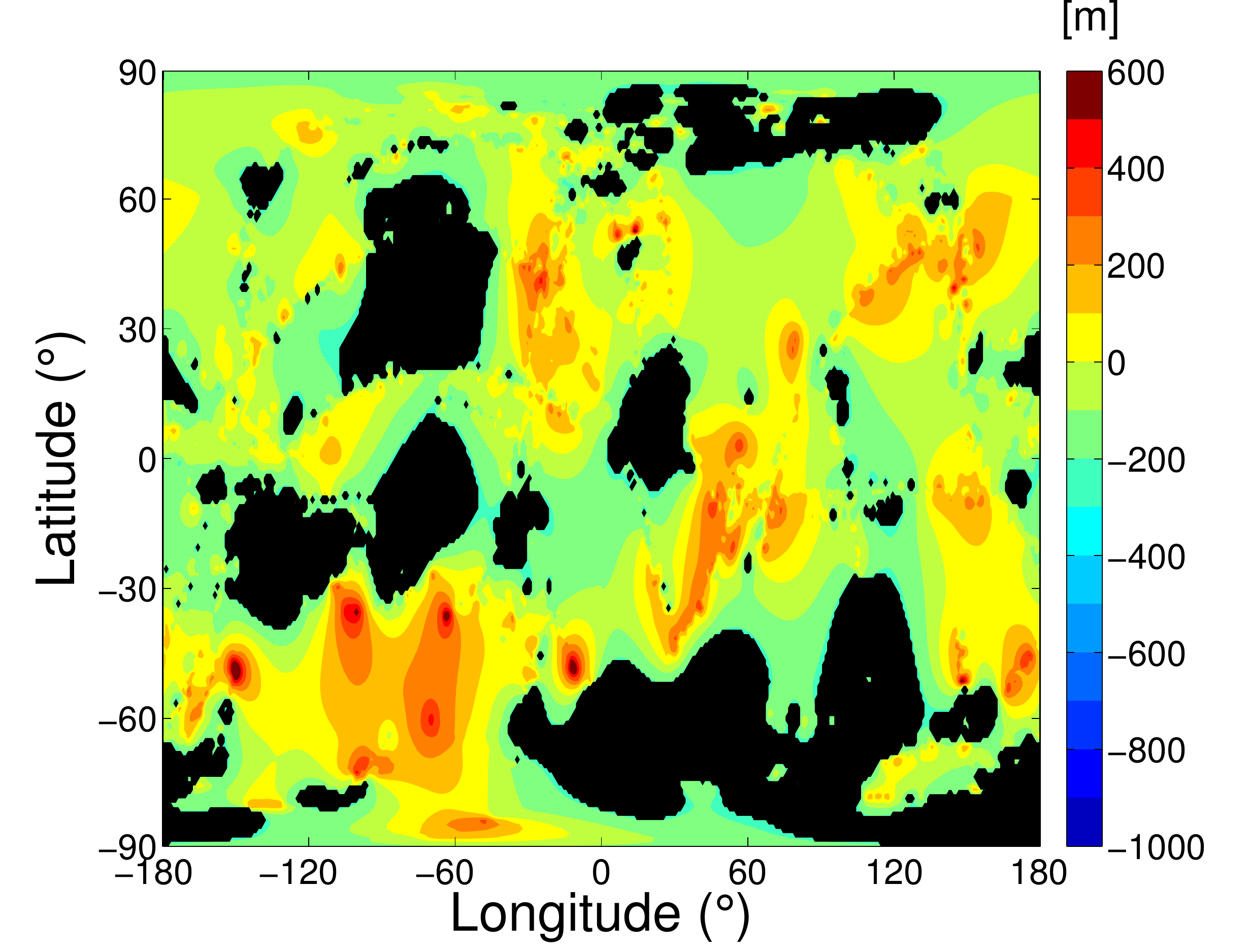}\\
\includegraphics[width=20pc]{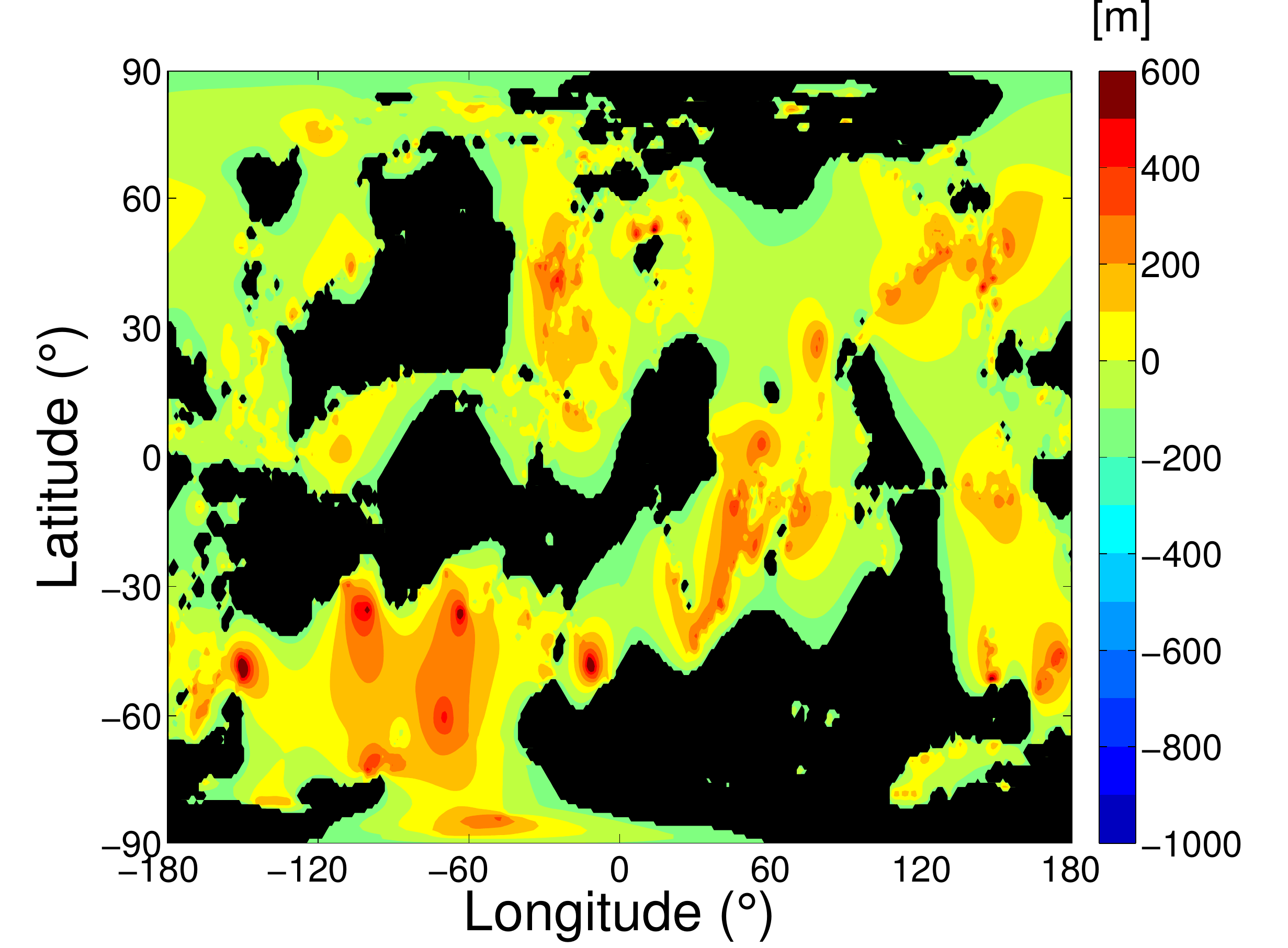}\\
\includegraphics[width=20pc]{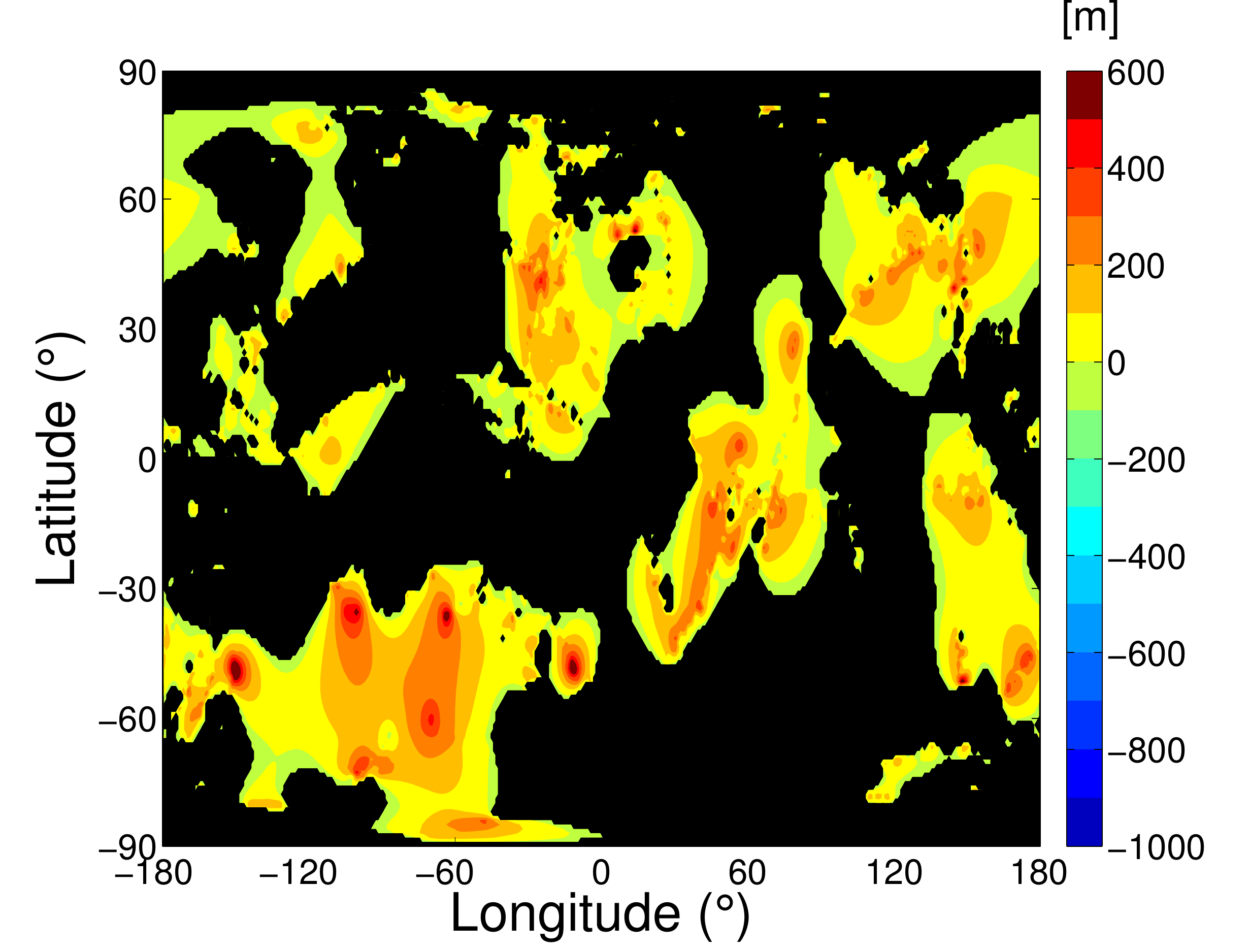}
\caption{Distribution of liquid nitrogen coverage of Titan for a condensation of 0.3 bar (top), 0.5 bar (middle) and 1 bar (down) of the atmosphere. The coverage is shown in black and corresponds to 21$\%$, 34$\%$ and 58$\%$ of Titan's surface, respectively. The topographic map from \cite{lorenz13} was used taking into account the geoid of \cite{iess10}.}
\label{fig_topo}
\end{center}
\end{figure}

In all the simulations, condensation happens at latitudes higher than 50$^\circ$ corresponding to cold traps. In the cases of strong condensation (early climates or surface albedo at 0.4-0.5), the polar nitrogen seas have a depth up to 1 km (see Fig. \ref{fig_n2_ts}). Such depths cannot be maintained, and liquid nitrogen would naturally flow to lower latitudes where it would evaporate. 
However, it is unlikely that Titan was fully covered by liquid nitrogen, because of the topography and the polar flattening \citep{zebker09}. It would require an amount of condensed nitrogen larger than present-day atmospheric nitrogen amount. Moreover, if Titan was fully covered by liquid, Saturn's tides would have produced a strong dissipation incompatible with the current eccentricity of Titan's orbit \citep{sagan82}. A non global ocean or disconnected seas could be acceptable \citep{dermott95}.
According to the results by \cite{larsson13} for a paleo-ocean of methane with current Titan's topography, a condensation of 0.3 bar of the atmosphere (e.g. a mean depth of around 30 m of liquid nitrogen) would  correspond to a liquid coverage of around 25$\%$ and would produce a southern polar ocean flooding the equator. These calculations are based on the relatively flat spherical harmonic topography of \cite{zebker09} and are likely optimistic for the equatorial flooding. We have reproduced them with the more detailed topographic map of \cite{lorenz13}. Fig. \ref{fig_topo} shows the topography and the coverage of liquid nitrogen for different total amounts of condensed nitrogen. For these calculations, we consider that liquid nitrogen is stable everywhere and first fill the lowest elevated areas. In reality, liquid nitrogen lakes at low latitude should evaporate if they are not connected to polar lakes. According to Fig. \ref{fig_topo}, liquid nitrogen should therefore remain confined in high latitudes for a condensation of 0.3 bar of the atmosphere. For a condensation of 0.5 bar, the southern polar sea should flood a small portion of low latitudes. For a condensation of one bar or more, liquid nitrogen covers 58$\%$ of Titan's surface, including most of the equatorial regions, and the polar seas are connected.
According to our model, the second case (i.e. condensation of 0.5 bar and moderate flooding) is produced at 3 Ga with an albedo of 0.3 and at 1.5 Ga with an albedo of 0.4 and non-radiative clouds. The third case (i.e. condensation of 1 bar and strong flooding) is produced at 3-4 Ga with an albedo at 0.5 and non-radiative clouds.

If the inventory of nitrogen (in the atmosphere and at the surface) has not significantly changed during Titan's history, a strong flooding is unlikely during the last 2 billions of years. However, a moderate flooding could have occurred. It would have triggered a cooling of low latitudes and a wetter climate, as discussed below.

\subsection{Simulation of wet climates}
\begin{figure}
\begin{center}
\includegraphics[width=20pc]{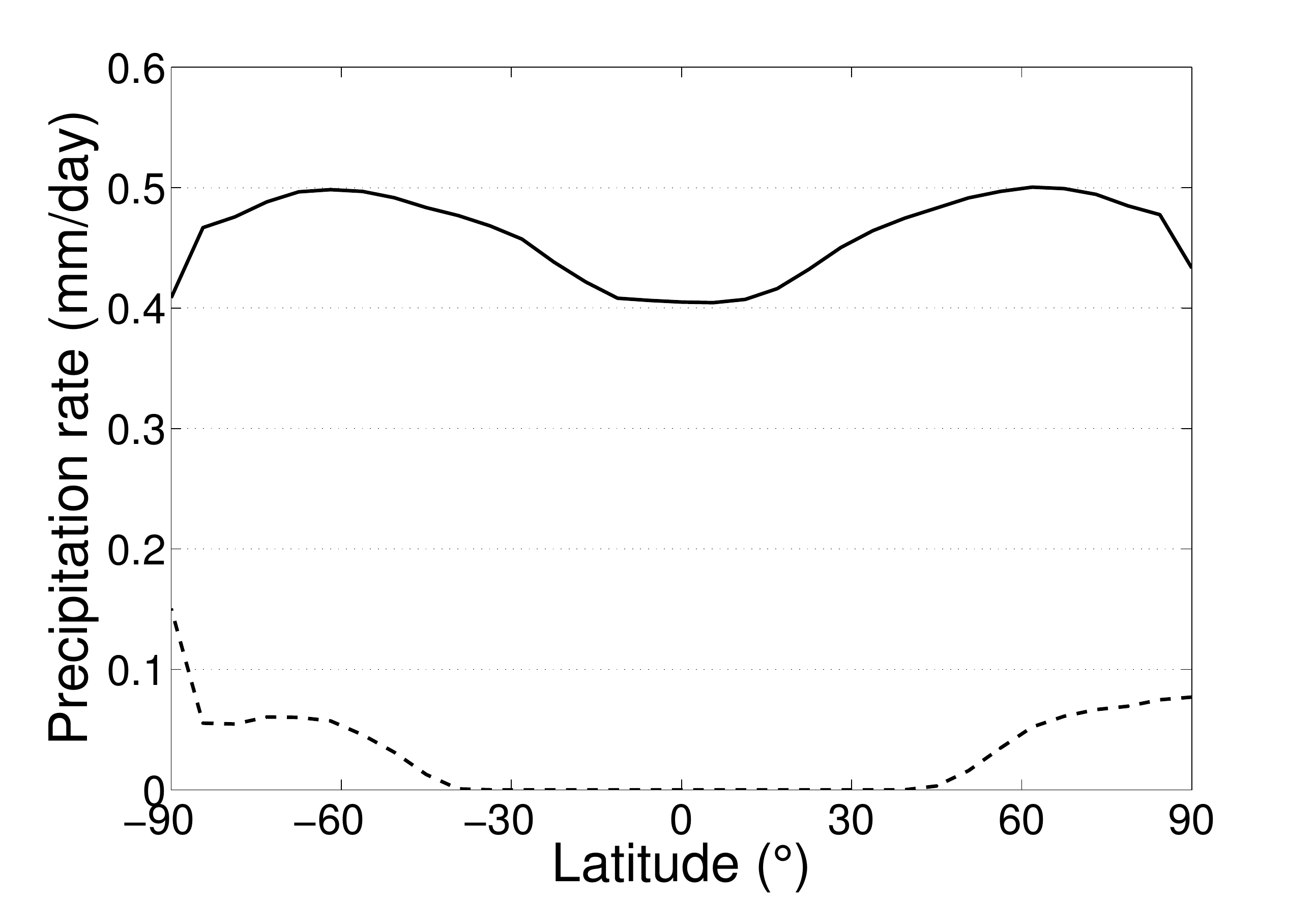}\\
\caption{Zonally and annually averaged precipitation at 4 Ga with horizontal diffusion (filled line) and without horizontal diffusion for liquid nitrogen (dashed line).}
\label{fig_rain_lat}
\end{center}
\end{figure}

We consider here the case of an extensive flooding (i.e. condensation of 1 bar of the atmosphere), with liquid nitrogen covering most of Titan's surface. 
To mimic this scenario, we ran a simulation at 4 Ga with a surface albedo at 0.3 and with horizontal diffusion of liquid nitrogen. Fig. \ref{fig_rain_cond_diff} shows the depth and the accumulation of liquid nitrogen on the surface during one Titan year. With diffusion, the liquid nitrogen reaches lower latitudes where it evaporates. 
Under these conditions, the surface temperature is fixed at 76 K all over Titan by liquid vapor equilibrium. Precipitation reaches the ground everywhere and have small latitudinal variations (see Fig. \ref{fig_rain_lat}).

\begin{figure}
\begin{center}
\includegraphics[width=20pc]{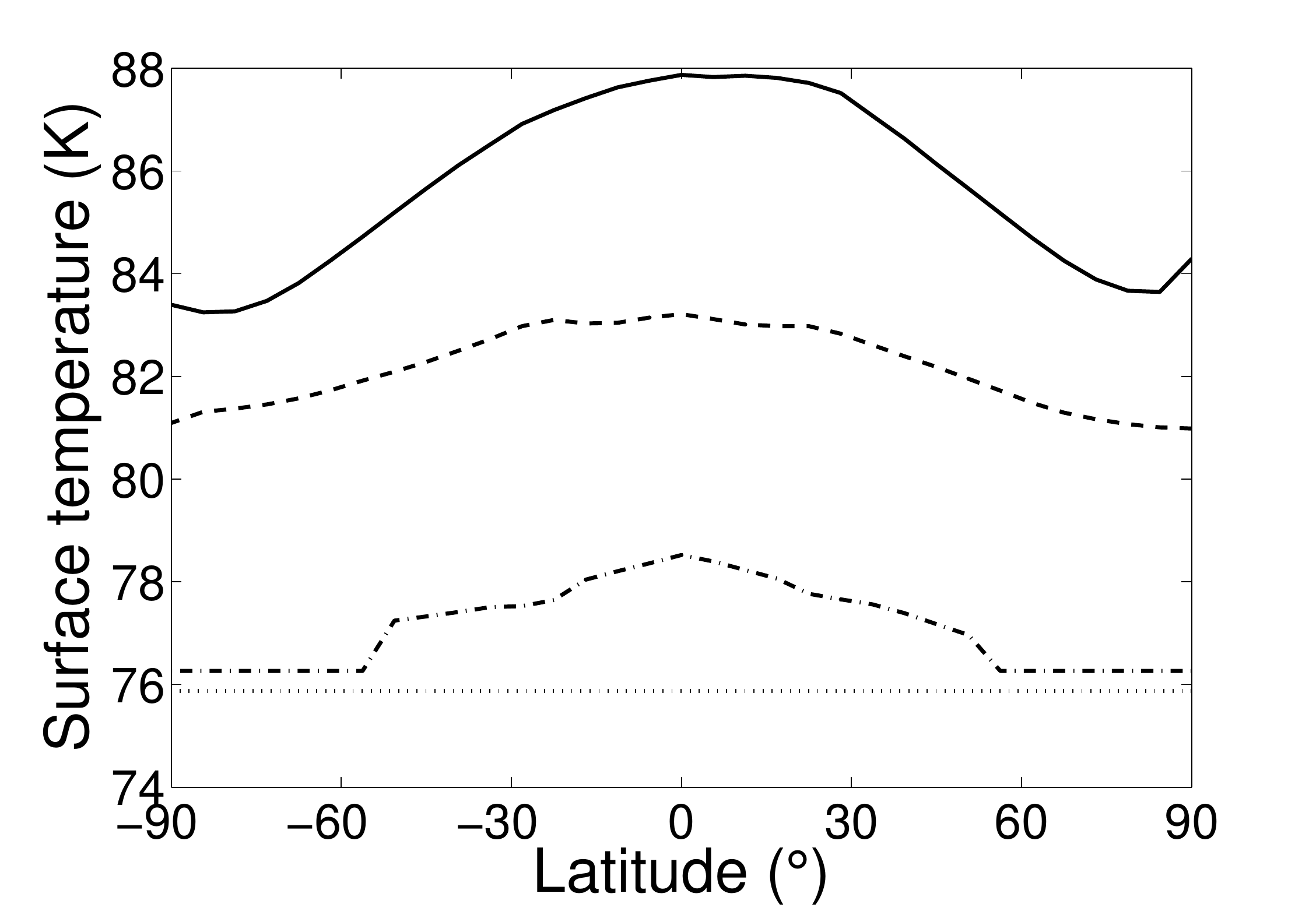}\\
\caption{Zonally and annually averaged surface temperature at present with an albedo at 0.2 (fill line), at 1 Ga with an albedo at 0.3 (dashed line), and at 4 Ga with an albedo at 0.3 (dashed-dotted line). The dotted line corresponds to the latter simulation with horizontal diffusion for liquid nitrogen.}
\label{fig_tsurf_lat}
\end{center}
\end{figure}

Fig. \ref{fig_tsurf_lat} shows the latitudinal profile of mean surface temperature for different cases (future, recent and early climates, and early climate with diffusion of liquid nitrogen). The equator-pole surface temperature gradient is around 4.5 K for future climates and around  2-2.5 K for past climates without horizontal diffusion. This temperature gradient of 2-2.5 K is also obtained with any values of albedo and solar flux providing there is a permanent polar condensation.
With horizontal diffusion, the equator-pole temperature gradient is null and polar surface temperatures are slightly reduced, but the impact on the surface pressure remains very small.

\begin{figure}
\begin{center}
\includegraphics[width=20pc]{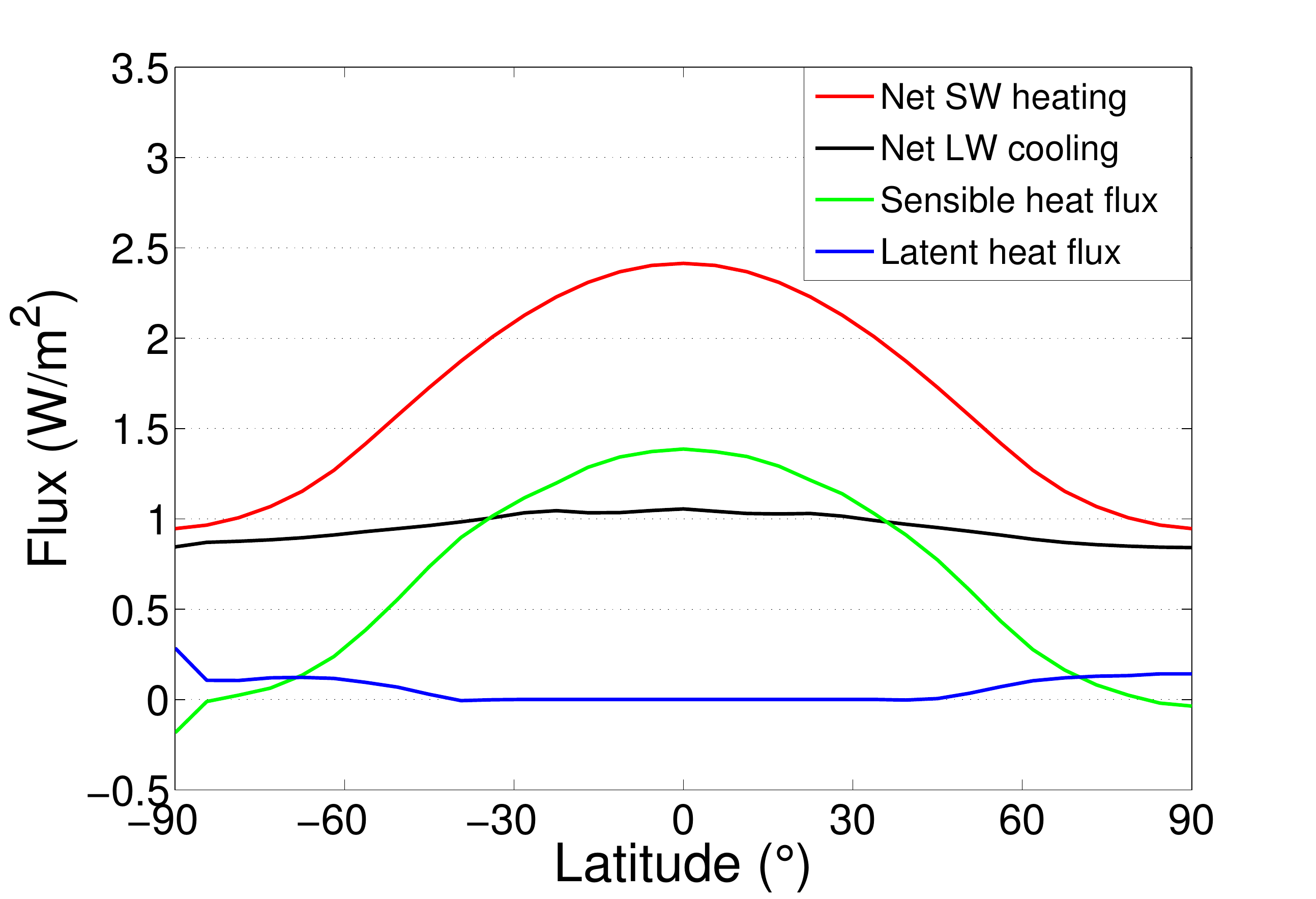}\\
\includegraphics[width=20pc]{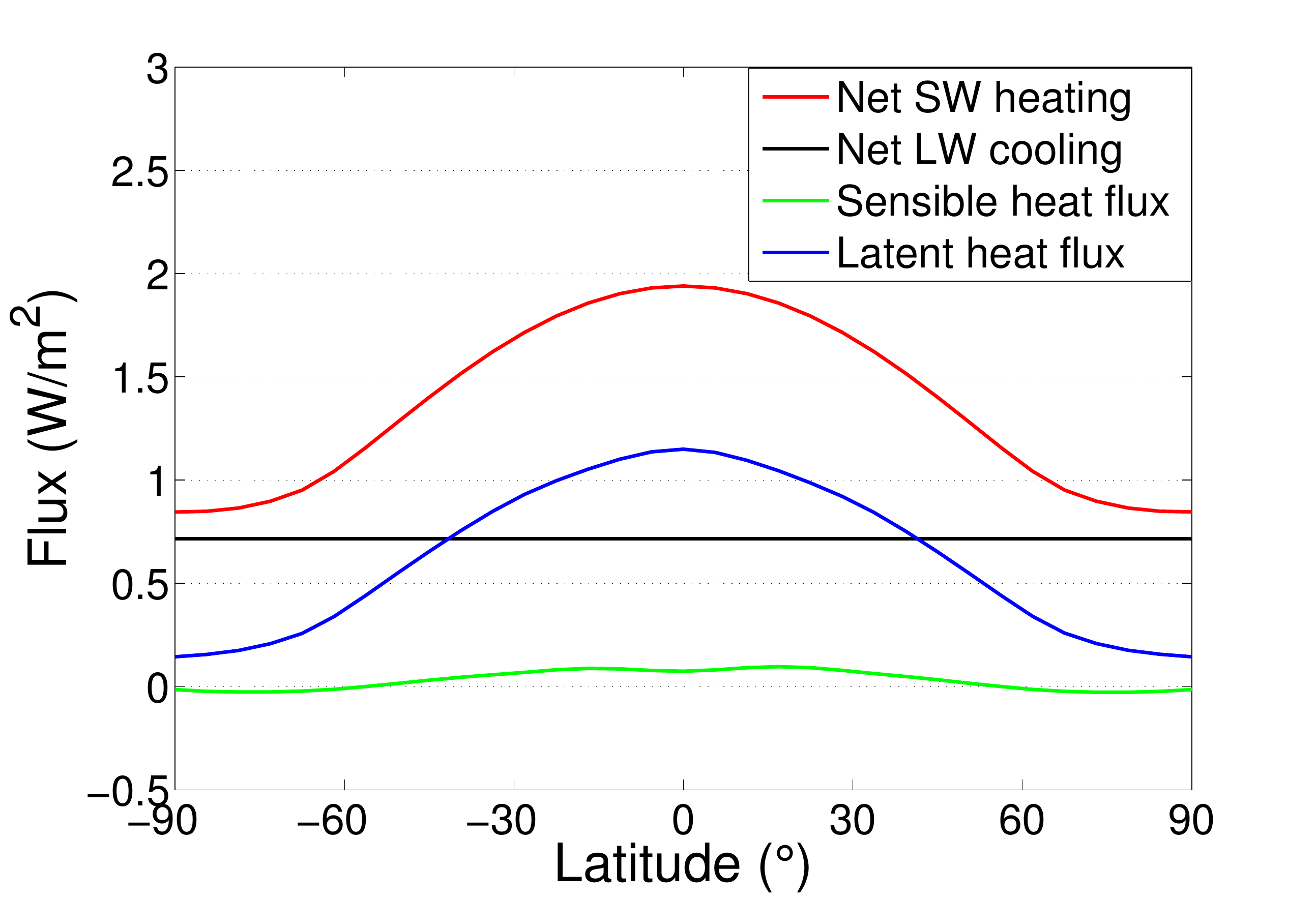}
\caption{Zonally and annually averaged fluxes at the surface: the net shortwave solar heating (in red), the net infrared cooling (in black), the sensible and latent heat flux from surface to atmosphere
(in green and blue respectively).
Top panel is the simulation at 1 Ga with the surface albedo at 0.3. Bottom panel is the simulation at 4 Ga with the surface albedo at 0.3 and  with horizontal diffusion for liquid nitrogen.}
\label{fig_flux}
\end{center}
\end{figure}

Fig. \ref{fig_flux} shows the different fluxes at the surface (net shortwave heating, net longwave cooling, sensible heat flux and latent heat flux) for a simulation at 1 Ga with no nitrogen diffusion and a simulation at 4 Ga with nitrogen diffusion. In the first case, there is a strong sensible heat flux (around 0.9 W/m$^2$ in average), while the latent heat flux is very weak (around 0.03 W/m$^2$ in average). 
In the second case, most of the sensible heat flux is converted into latent heat flux (around 0.8 W/m$^2$) corresponding to a high precipitation rate (around 0.46 mm/day in average). 
The nitrogen cycle is therefore strengthened when liquid nitrogen floods low latitudes, producing a wet climate with enhanced precipitation (see Fig. \ref{fig_rain_lat}).
For a moderate flooding (e.g. a condensation of 0.5 bar), the climate should be intermediate between this wet climate and the dry climate described before. All the surface would not be covered by liquid nitrogen, but the air temperature should be cooled, even at low latitudes. We therefore expect that drizzle would reach the land surface in this case, at least at mid and high latitudes.

\section{Possible implications for erosion, and crustal exchanges}

\subsection{Possible impact on erosion and the age of the surface}
According to our model, the nitrogen cycle for a methane depleted atmosphere may be stronger than the present methane cycle on Titan. Hence, more rain can be produced under the nitrogen cycle, in particular when polar nitrogen seas flood low latitudes, producing a wet climate. When the surface is fully covered by liquid nitrogen, 0.5 mm/day of liquid nitrogen falls globally. This is larger than what is expected with present methane rain \citep{schneider12}.
They are two reasons for this difference. 
Firstly, nitrogen is far more volatile than methane. A variation of temperature produces condensation larger by one order of magnitude for nitrogen than for methane.
Secondly, the solar insolation absorbed by the surface is higher (around 5 times higher) for Titan depleted in methane (i.e. no absorption by haze and hydrocarbons), enhancing the sensible heat flux and thus evaporation and precipitation. The sensible heat flux is around 1 W/m$^2$ for a dry climate without methane and around 0.23 W/m$^2$ for present Titan with no methane cycle \citep{charnay12}. 
The maximal latent heat is therefore limited by these values, up to 1 W/m$^2$ for a nitrogen cycle and up to 0.2-0.3 W/m$^2$  for a methane cycle (consistent with the value around 0.3 W/m$^2$ for the moist case in \cite{mitchell12}).
This difference leads to a maximal globally averaged precipitation rate of around 0.5 mm/day for nitrogen and 0.1 mm/day for methane with current atmosphere (we consider that all the sensible heat flux is converted into latent heat flux, as in the bottom panel in Fig. \ref{fig_flux}).

Past nitrogen rain may therefore have been stronger on average than methane rain. However, our model predicts that nitrogen precipitation would have been a permanent light drizzle everywhere rather than rare violent downpours as for present equatorial methane rain \citep{turtle11a, mitchell11, schneider12}, which are associated to a high accumulation of liquid methane at the surface of the order of 100 kg in a few hours \citep{hueso06, barth07}. For a pure atmosphere, convective clouds can only form if there is supersaturation. This is the case for CO$_{2}$ clouds on the present Mars \citep{colaprete08}. We expect no supersaturation or only a weak supersaturation in Titan's troposphere depleted of methane (see section 3 'Effect of nitrogen clouds'). Convective clouds, associated with more vigorous precipitation could therefore form, but without any observational constraints, this remains speculative. We therefore consider only drizzle on methane-depleted Titan, keeping in mind that precipitation could be more violent than predicted by our GCM. 

\begin{figure}
\begin{center}
\includegraphics[width=20pc]{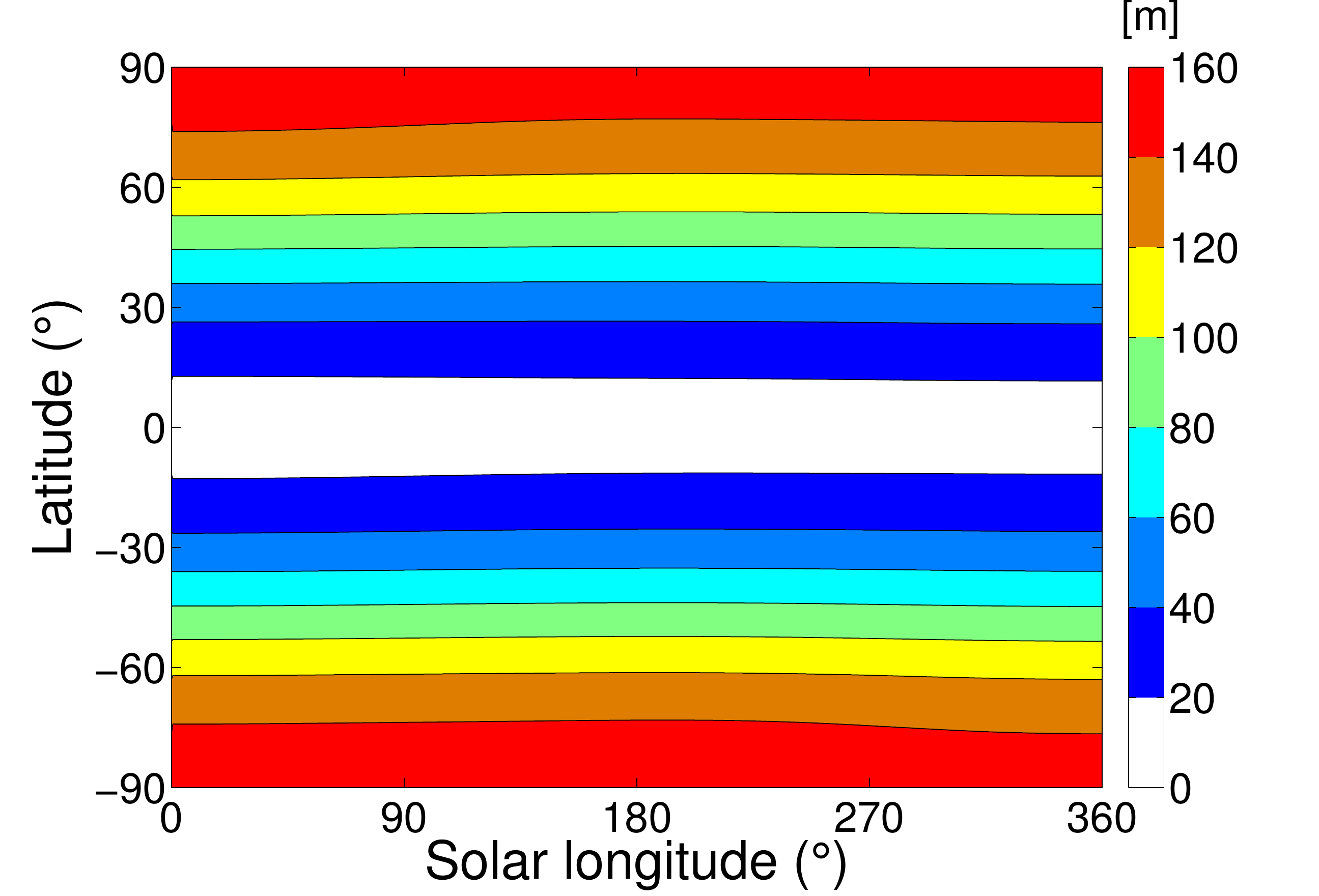}\\
\includegraphics[width=20pc]{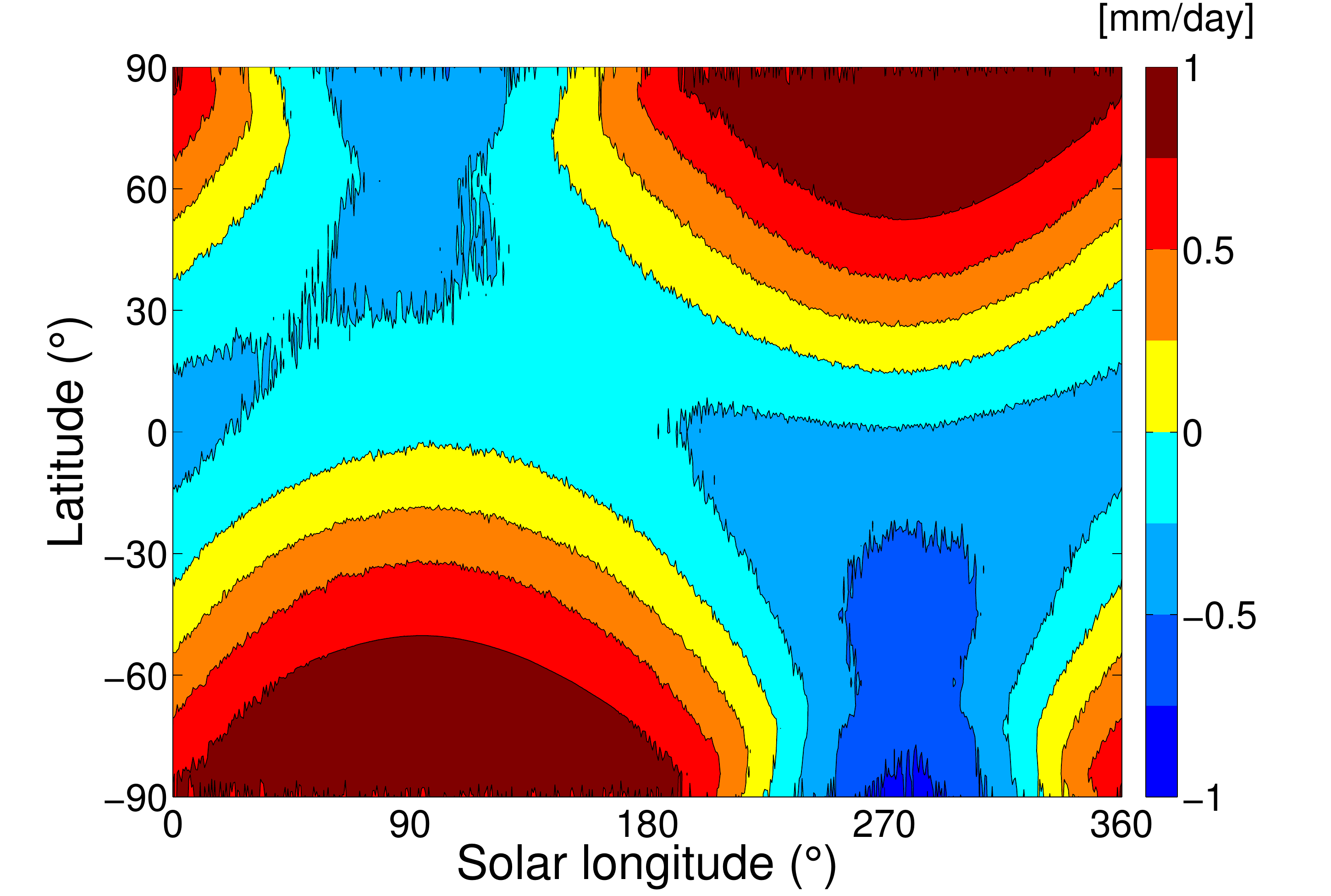}
\caption{
Zonally averaged depth of liquid nitrogen (in m) on the surface (top) and accumulation of liquid nitrogen (down, same as Fig. \ref{fig_rain_cond}) depending on latitude and solar longitude. Simulation performed at 4 Ga with the surface albedo at 0.3 and with horizontal diffusion for liquid nitrogen and with non-radiative clouds.}
\label{fig_rain_cond_diff}
\end{center}
\end{figure}

A drizzle is far less efficient to transport sediment than rainstorms. But over a long period of time it may erode the surface. Moreover, since the density of liquid nitrogen (810 kg/m$^3$) is close to the density of water ice (920 kg/m$^3$), unlike methane (450 kg/m$^3$), nitrogen flows would have been very efficient to transport ice rocks and then to erode Titan's surface. The precipitation rate threshold to transport sediment is proportional to $(\rho_{sed}/\rho_{liq}-1)^{\frac{3}{2}}$, with  $\rho_{sed}$ the sediment density and $\rho_{liq}$ the liquid density \cite[]{perron06}. 
This threshold to transport water ice rock on Titan is thus around 21 times lower for liquid nitrogen than for liquid methane.
This higher erosion ability, added to the high mean precipitation rate and the nitrogen flows produced by a strong condensation of the atmosphere, therefore makes the nitrogen cycle potentially efficient to erode Titan' surface.
The transport threshold is estimated to be around 0.5 mm/hour to 15 mm/hour for methane on Titan\cite[]{perron06}. It is thus around 0.6 mm/day to 17 mm/day for liquid nitrogen. The lowest values are compatible with accumulation rates from our GCM for a strong condensation of the atmosphere, reaching up to 1 mm/day at high latitudes and up to 0.5 m/s at mid latitudes (e.g. 40$^\circ$) (see Fig. \ref{fig_rain_cond_diff}). Moreover, our GCM calculates average precipitation and surface condensation over large areas. At smaller scales and taking into account topography, we could expect to have larger accumulation rates locally. We therefore conclude that nitrogen surface condensation and rain may erode the surface and carve valleys at mid and high latitudes. At low latitudes, the evaporation of nitrogen at the surface (see Fig. \ref{fig_rain_cond_diff}) should decrease its erosional potential. We therefore conclude that nitrogen rain is unlikely to erode low latitudes, unless there is supersaturation in the atmosphere, producing episodic strong precipitation by convective clouds.

The analysis of crater distribution indicate a lower amount of craters at low elevation
\citep{neish14}. This could be explained by the presence of extensive wetlands of liquid hydrocarbons at low elevation during the last hundred million years, consistent with the age estimations for methane \citep{neish14}. This asymmetry in crater distribution may similarly be explained by a nitrogen flooding and wetting of low elevations in the case of a strong condensation. The flooding map we produced (Fig \ref{fig_topo} middle) is consistent with the crater map of \cite{neish14}. Craters have been detected in Xanadu which is flooded in Fig \ref{fig_topo}. However, this region is not connected to polar lakes, thus liquid nitrogen should not remain there. This climate with wetlands of liquid nitrogen has the advantage to be stable for hundreds of millions of years but could only have occurred before 1 Ga.

The surface of Titan is young, between 200 million and 1 billion years, according to impact crater counting \citep{wood10,neish12}. This age is similar to the onset of methane outgassing in the model of \cite{tobie06}. 
If Titan was depleted in methane before this time, the nitrogen cycle could have eroded the surface until the outgassing of methane, which was maybe less efficient to erode all the surface. The measured age would be directly linked to the end of the nitrogen cycle. 
In addition, if a strong condensation occured, nitrogen rain and seas could have produced some of the shorelines and valley networks observed at mid and high latitude \citep{moore10}. At this stage, this remains a speculation, and most Titan's fluvial features have more likely been produced by methane rain on present or past Titan. Nevertheless, it is interesting to reveal a new process for erosion.

\subsection{Crustal exchanges and consequences for  Titan's shape and methane outgassing}

During periods of permanent nitrogen condensation, a significant fraction of liquid nitrogen may also have infiltrated in the crust, as it has been suggested for liquid methane and ethane \citep{kossacki96,hayes08,choukroun12}.   A crust with about 10$\%$ porosity could easily accommodate the equivalent of 0.5 bar of nitrogen as liquid.  A global "aquifer" consisting of liquid nitrogen may therefore have been present on Titan, especially during the earliest epoch.  Although not explicitly included in our simulations, the presence of such a subsurface reservoir may have strengthened the nitrogen cycle, increasing the precipitation rate at lower latitudes, as shown for simulations with horizontal diffusion of liquid nitrogen. 

\begin{figure}
\begin{center}
\noindent\includegraphics[width=20pc]{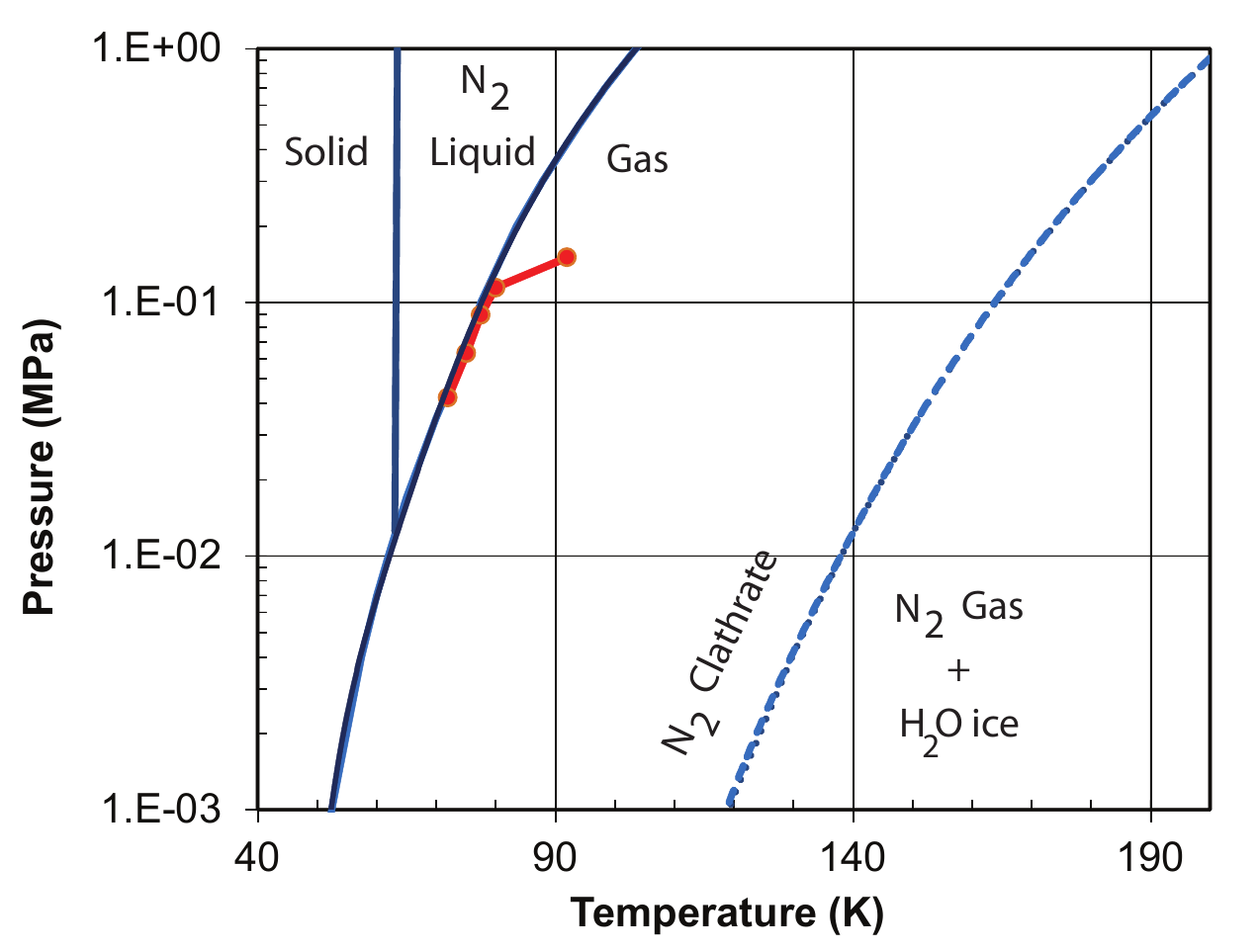}
\caption{Phase diagram of N2 (solid lines) and N2-clathrate (dash lines). The red
points correspond to first four values in Table 2 except for the higher pressure one that
corresponds to present conditions on Titan’s surface. It shows that the surface
conditions are controlled by the liquid-vapor stability line. It also shows that N2
clathrate are very stable under Titan's surface conditions.}
\label{fig_N2-phasediagram}
\end{center}
\end{figure}

The presence of liquid nitrogen in crustal pores, down to depth of about 1 km, may also affect the chemical evolution of the crustal materials, especially in the polar regions where liquid nitrogen accumulates in our models. Similarly to what has been predicted by \cite{choukroun12} for ethane, nitrogen may be progressively sequestrated in the form of clathrate hydrate, either by transformation of crustal ice into clathrate hydrates (see figure \ref{fig_N2-phasediagram} for the stability of N$_2$ clathrates) or by substitution between methane and nitrogen molecules in clathrate hydrate structures.  The methane released by this mechanism could either stay in the crust or go into the atmosphere producing a methane outgassing. Yet, the surface would be too cold (likely lower than 81 K at poles) to have methane in a liquid state. It would concentrate in the polar regions, frozen or dissolved in the nitrogen lakes until the surface is warmed by, for instance, a major impact or the increase of solar irradiance. This deglaciation could have suddenly released large amount of methane in the atmosphere, switching from a nitrogen cycle to the present methane cycle.

Another consequence of the nitrogen sequestration in the form of clathrate would be an increase of crust density. Indeed, water ice and/or methane clathrate has density of about 930-940 kg.m$^{-3}$, whereas nitrogen-dominated clathrate has a density of about 1000 kg.m$^{-3}$ \citep{sloan98}. The sequestration of nitrogen should therefore lead to a progressive increase of crust density. By considering ethane, which has similar consequences on density, \cite{choukroun12} showed that the observed excess polar flattening of 270 m compared to the hydrostatic reference case may be explained by accumulation of dense clathrate at the poles.  They estimated that 300-1200 million years at the present ethane production rate were required to produce the observed flattening. N$_2$ having a molecular mass similar to ethane,  a sequestration of 0.25-1 bar of nitrogen in the clathrate crust in the polar regions would produce a similar effect.  This implies that a significant  fraction of the primordial nitrogen may be trapped in the polar crust, the observed 1.47 bar in today atmosphere would therefore be the remnant of a more massive nitrogen inventory.  However, as the atmospheric pressure is limited by the saturation pressure, most of the excess nitrogen would be in the form of liquid at the surface or in the subsurface, and this would not change significantly the results of our model. When Titan's climate switched to the present methane cycle, all remaining liquid nitrogen would have been released to the atmosphere as gaseous nitrogen, whereas nitrogen incorporated in clathrate structure would remain trapped in the crust.

\section{Conclusion}

This study constitutes a detailed analysis of a pure nitrogen atmosphere on Titan. We have shown that if Titan was methane depleted during some periods of its history, a strong nitrogen cycle with nitrogen seas, clouds and precipitation could have happened. Titan could fall into this state in the future. A pure nitrogen atmosphere is more complex than it appears. Our analysis of nitrogen clouds show that they would likely be rather optically thin, not able to warm efficiently the surface. One of the main features is the stability of liquid nitrogen at the poles and its very short lifetime at the equator where it evaporates quickly. Therefore, this 3D modeling validates the previous 1D study by \cite{lorenz97} but also reveals new major implications.

Concerning Titan's paleo-climates during the last two billion years, the present study suggests that a nitrogen cycle could have contributed to the erosion of Titan's surface. It may explain the surface age and some of the valley networks observed at present at mid and high latitudes, although the methane cycle remains the most likely explanation. The polar nitrogen seas could be at the origin of the excess flattening, because liquid nitrogen in contact with water ice forms nitrogen clathrates that are denser. This transformation being more efficient at polar areas than at low latitudes where nitrogen evaporates, it induces a lateral density change that can explain the observed excess flattening. Experimental data are required to investigate the timescales involved in the formation of nitrogen clathrates in the crust. If the crust was made of methane clathrates, not only nitrogen would replace methane in clathrates but also methane would be outgassed. Simulating such a process and its climatic consequences requires 3D models similar to our GCM, but including methane, hydrogen and haze acting as CCN.

Concerning the early history of Titan, we showed that, under the faint Sun, a substantial part of Titan's atmosphere could have condensed forming deep nitrogen seas. A collapse into a Triton-like solid nitrogen frozen state would have required a surface albedo higher than 0.65. With a surface albedo of 0.8 or more, this state would be stable even today.
However, a surface albedo higher than 0.65 seems unlikely considering the albedo of icy satellites. The pressure is likely to have remained above the triple point of nitrogen. Thus, our analysis gives constraints on the minimal atmospheric pressure for the past of Titan. Such constraints provide clues to understand the formation, evolution and escape of the atmosphere during Titan's history.

Finally, even if a paleo-nitrogen cycle remains speculative, its implications for Titan's geology and evolution are potentially so profound that it deserves attention. A natural perspective of future research will be to use the 3D GCM, including methane, hydrogen and haze, to investigate Titan's past climates with different amounts of methane. Such studies have already been performed with 1D models, particularly to simulate the evolution of Titan's atmosphere over a methane-ethane ocean \citep{lunine89,mckay93, lorenz99}. A new look on these issues with a 3D GCM could bring accuracy and clear benefits. In particular, it will be interesting to simulate the stability of Titan's climate with a small amount of methane. This will allow to analyze the impact of a methane outgassing and the change from a nitrogen to a methane cycle.
\\
\\
Acknowledgments: CS acknowledges support by the NASA Astrobiology Institute program.
\\
\\

\bibliographystyle{elsarticle-harv}




%% \bibitem must have one of the following forms:
%%   \bibitem[Jones et al.(1990)]{key}...
%%   \bibitem[Jones et al.(1990)Jones, Baker, and Williams]{key}...
%%   \bibitem[Jones et al., 1990]{key}...
%%   \bibitem[\protect\citeauthoryear{Jones, Baker, and Williams}{Jones
%%       et al.}{1990}]{key}...
%%   \bibitem[\protect\citeauthoryear{Jones et al.}{1990}]{key}...
%%   \bibitem[\protect\astroncite{Jones et al.}{1990}]{key}...
%%   \bibitem[\protect\citename{Jones et al., }1990]{key}...
%%   \harvarditem[Jones et al.]{Jones, Baker, and Williams}{1990}{key}...
%%


% \end{thebibliography}

%% Enter Figures and Tables here:

\clearpage

\begin{table}
\begin{center}
\caption{Dependence on the CCN concentration for surface temperature, pressure, planetary albedo and cloud optical depth in visible. Simulations done with the present solar irradiance and a surface albedo of 0.3. The case with non-radiative cloud has been run with a CCN concentration of 10$^3$ kg$^{-1}$.\label{table1}}
\begin{tabular}{|c|cccc|}
\hline
\hline
CCN (kg$^{-1}$) & T$_s$ (K) & P$_s$ (bar) & Albedo & Cloud optical depth\\
\hline
Supersaturation& 88.3 & 1.47 & 0.48 & 0\\ 
Non-radiative cloud & 83.8 & 1.47 & 0.48 & 0 \\ 
10$^2$ & 85.1 & 1.47 & 0.51 & 0.9\\ 
10$^3$ & 87.3 & 1.47 & 0.55 &2.7\\ 
10$^4$ & 90.8 & 1.47 & 0.62 &6.9\\ 
10$^5$ & 91.6 & 1.47 & 0.71 &16\\ 
10$^6$ & 74.7 & 0.72 & 0.88 &75\\ 
10$^7$ & 63.9 & 0.14 & 0.9 &95\\ 
\hline
\end{tabular}
\end{center}
\end{table}

\begin{table}
\begin{center}
\caption{Surface pressure and temperature at 4 Ga depending on the surface albedo (non-radiative cloud).\label{table2}}
\begin{tabular}{|c|cc|}
\hline
\hline
Surface albedo & P$_s$ (bar) & T$_s$ (K)  \\
\hline
0.2 & 1.14 & 79.9\\
0.3 & 0.89 & 77.5\\
0.4 & 0.63 & 75.1\\
0.5 & 0.41 & 72.0\\
0.6 & 0.21 & 67.2\\
0.7 & 0.09 & 61.7\\
0.8 & 0.03 & 56.5\\

\hline
\end{tabular}
\end{center}
\end{table}

\end{document}